\documentclass[twocolumn,prb]{revtex4}
\usepackage{amsmath}
\usepackage{amssymb}
\usepackage{esint}
\usepackage{graphicx}

\makeatletter
\@ifundefined{textcolor}{}
{%
 \definecolor{BLACK}{gray}{0}
 \definecolor{WHITE}{gray}{1}
 \definecolor{RED}{rgb}{1,0,0}
 \definecolor{GREEN}{rgb}{0,1,0}
 \definecolor{BLUE}{rgb}{0,0,1}
 \definecolor{CYAN}{cmyk}{1,0,0,0}
 \definecolor{MAGENTA}{cmyk}{0,1,0,0}
 \definecolor{YELLOW}{cmyk}{0,0,1,0}
 }

\@ifundefined{definecolor}
 {\@ifundefined{definecolor}
 {\@ifundefined{definecolor}
 {\usepackage{color}}{}
}{}
}{}

\newcommand{\Ket}[1]{\vert  #1  \rangle}

\newcommand{\MatEl}[3]{\langle \, #1 \,\vert\,#2\,\vert\,#3\,\rangle}
\newcommand{\Amp}[2]{\langle  #1 \vert  #2  \rangle}

\renewcommand{\phi}{\varphi}
\renewcommand{\epsilon}{\varepsilon}
\renewcommand{\vec}[1]{{\bf #1}}

\begin{document}

\title{The theory of coherent dynamic nuclear polarization in quantum dots}
\author{Izhar Neder$^{1,4}$, Mark S. Rudner$^{2,3,4}$, Bertrand I. Halperin$^{4}$}
\affiliation{
$^{1}$ Raymond and Beverly Sackler School of Physics and Astronomy, Tel-Aviv University, Tel Aviv, 69978, Israel\\
$^{2}$ The Niels Bohr International Academy, Niels Bohr Institute, 2100 Copenhagen, Denmark\\ 
$^{3}$ Center for Quantum Devices, Niels Bohr Institute, 
2100 Copenhagen, Denmark\\
$^{4}$ Department of Physics, Harvard University, Cambridge, MA 02138, USA
}



\begin{abstract}
We consider the dynamic nuclear spin polarization (DNP) using two
electrons in a double quantum dot in presence of external magnetic
field and spin-orbit interaction, in various schemes of periodically
repeated sweeps through the S-T+ avoided crossing. By treating the
problem semi-classically, we find that generally the DNP have two
distinct contributions - a geometrical polarization and a dynamic
polarization, which have different dependence on the control parameters
such as the sweep rates and waiting times in each period. Both terms
show  non-trivial dependence on those control parameter.
We find that even for small spin-orbit term, the dynamical polarization
dominates the DNP in presence of a long waiting period near the S-T+ avoided crossing, 
of the order of the nuclear Larmor precession periods. A detailed
numerical analysis of a specific control regime can explain the oscillations
observed by Foletti et.~al.~in arXiv:0801.3613.
\end{abstract}
\maketitle

\section{Introduction}

In recent years, 
the interaction between 
electron 
and nuclear spins in quantum dots 
has received wide attention due to the system's potential as a platform for quantum information processing
\cite{Petta2005,Koppens2008,Greilich2006,Foletti2009,Nowack2007,Barthel2009,Loss1998,Levy2002, Taylor2005}.
In particular, this system is attractive due to the possibility of fast, local {\it electrical} control of electron spins: 
in quantum dots, electron spins can be manipulated using nearby electrostatic gates due to various forms of spin-orbit coupling\cite{Dresselhaus1955, Rashba1960, Bychkov1983}, 
including the ``spin blockade'' effect\cite{OnoScience}, which 
couple the spin and orbital degrees of freedom of the electron. 
However, the interaction between the electron spins
and the nuclear spins of the host lattice presents a significant source of dephasing in electron spin qubits based on quantum dots in GaAs and similar materials\cite{Taylor2005,Taylor2007,HansonRMP,Merkulov2002,Erlingsson2002,Johnson2005,Sousa2003,Koppens2005}.
Thus for the successful application of these devices it is necessary to develop a detailed understanding of the quantum dynamics of this coupled many-body spin system, as well as means to control and mitigate the unwanted processes which lead to electron spin decoherence.

With this motivation, an intense experimental effort is underway to 
control and detect electron spins ever faster, and with improving accuracy\cite{Barthel2009,ourselves}.
In addition, many groups are exploring ways of fighting nuclear spin induced electron spin decoherence by using fast electron spin control to dynamically polarize and manipulate the nuclear spin bath via the hyperfine interaction
{\cite{Klauser2006,Ramon2007,Reilly2008B,Foletti2009,Ribeiro2009,Gullans2010,Gullans2012,BluhmFeedback,Yao2011}}. 
In particular, several recent experiments demonstrated that dynamical nuclear polarization (DNP) can be used to extend the electron spin coherence time\cite{BluhmFeedback}.
Furthermore, in Ref.~\onlinecite{Foletti2009} the authors used the controlled Overhauser-Zeeman field resulting from inhomogeneous DNP in a double quantum dot to manipulate the electron spin state, thus gaining 
universal control of the electron
spin qubit dynamics. 

Along with the motivation coming from potential applications, the system of coupled electron and nuclear spins in quantum dots is interesting from a fundamental point of view, due to the fact that the nuclear spins are relatively well isolated from their environment.
Nuclear spin coherence times may reach hundreds of microseconds to a millisecond\cite{Takahashi2011,Makhonin2011}, which is much longer than the nanosecond timescales associated with electron spin dynamics.
Thus the electron spin interacts with a ``bath'' in which long-lived 
quantum coherence may play an important role. 
For example, coherence in the nuclear spin state can lead to ``superradiance''-like effects\cite{CiracSuperradiance}.
Coherent Larmor precession of nuclear spins due to an applied magnetic field may also lead to interesting non-Markovian effects on the electron spin evolution\cite{Greilich2006, Greilich2007, ourselves, Cywinski2009, Cywinski2009B, Rudner2010, Neder2011, Glazov2012}.


In this paper, we focus our attention on the roles of nuclear spin coherence and of electronic spin-orbital coupling on DNP generated via fast electron spin manipulation.
Our main motivation is a recent experiment by Foletti et al., in which DNP was produced by repeatedly modulating the parameters of a GaAs two-electron double quantum dot, in the presence of an external magnetic field\cite{Foletti2008}. 
Strikingly, the experiment revealed a significant enhancement of DNP for cycle times which matched integer multiples of the Larmor precession periods of the Ga and As nuclei. 
In addition, the {\it sign} of the DNP, i.e.~the direction of nuclear polarization build-up, parallel or opposite to the applied field, 
was also reversed at the commensuration points.

Why are these results remarkable?
First, the observation that DNP is sensitive to the Larmor periods of individual nuclear spin species indicates that the system is somehow sensitive to the absolute phase of nuclear precession, in the laboratory frame.
What feature in the system can provide a reference relative to which such precession should be measured?
Second, how is it possible for the direction of DNP to reverse itself?
In the absence of spin-orbit coupling, conservation of angular momentum provides a selection rule which requires every electron spin flip from down to up to be accompanied by a compensating nuclear spin flip from up to down (and vice-versa).
Simple arguments based on this selection rule yield a prediction for the expected sign of DNP, which is violated here.
As we will discuss in detail below (see also Ref.~\onlinecite{Rudner2010}), the combined effects of electronic spin-orbit coupling and long-lived nuclear spin coherence can resolve these questions.

Here we show 
that these key principles, which underlie the model studied in Ref.~\onlinecite{Rudner2010}, can be extended and adapted to more realistic models of 
experiments in GaAs double quantum dots. 
In particular, we take into account the presence of three distinct nuclear spin species, $^{69}$Ga, $^{71}$Ga, and $^{75}$As, each with its own characteristic gyromagnetic ratio.
We emphasize the important similarities as well as new effects which arise in the multispecies case, as compared with a 
single species model\cite{Rudner2010}.

Dynamical nuclear polarization results from back-action of the electrons on the nuclei.
This process can be understood intuitively in terms of nuclear precession around the total field produced by the vector sum of the external field and the hyperfine (Knight) field produced by the electrons on the nuclei. 
Transverse components of electron spin relative to the external field tilt the nuclear precession axis, causing the projection of nuclear polarization on the external field axis to change with time. 
Employing a semiclassical approach in which the polarization of each nuclear species is treated as a classical (time-dependent) vector\cite{Erlingsson2004,Chen:semiclass,Al_Hassanieh2006,Brataas2011,Brataas2012, Gullans2010,Gullans2012}, 
we study specific experimentally relevant protocols.
In particular, we focus on a protocol similar to the one used in the experiment
in Ref.~\onlinecite{Foletti2008}.
Numerical results show an oscillatory dependence of the nuclear polarization rate on the cycle time, which resembles the experimental results. 

The oscillations result from subtle correlations in the electron spin (Knight field) evolution between successive sweeps, which appear in addition to the trivial correlations that would be expected in the absence of coupling to nuclei. The additional correlations arise because the electron spin evolution over a single sweep depends on the direction of the Overhauser field, which changes between sweeps due to nuclear Larmor precession.

In addition to numerical investigation, we develop a simple analytical treatment in which we separate the DNP production rate into two contributions 
which depend in very different ways on the parameters of the system and on its evolution. 
Using this separation, 
we estimate the DNP production rates from straightforward energy conservation arguments.
Both the numerical and the approximate analytical approaches are widely applicable, and may be 
be used to predict the outcomes of a variety of future DNP experiments.

In this work, we consider DNP starting from a high-temperature equilibrium ensemble of initial nuclear spin states, as is typically realized in experiments.
Because the changes in the nuclear polarizations 
generally depend in a complicated manner on the precise initial nuclear state, the most meaningful quantities that we can calculate are those which are, 
in some way, averaged over the initial ensemble.
In practice, the averaging is performed by selecting a representative set of initial states from the specified ensemble, and following separately the time evolution of each one.

For the case of the thermal (completely mixed) distribution, we are quite free in terms of the types of initial nuclear spin states that we choose.
Within the semi-classical treatment that we employ, 
the nuclear state at any given time is described by specifying the orientation of each individual nucleus. More economically, individual spins with similar hyperfine coupling to the electron spins may be grouped together into larger composite spins; the nuclear state is then described in terms of the orientations of these composite spins. 
We then average the results with respect to the initial orientations of the spins.

In this paper, 
we calculate quantities such as the average rates of change of the nuclear polarizations and of the related Overhauser fields, starting from the random distribution of initial states, over a limited number of DNP cycles (typically of the order of a few thousand). Over the length of our calculation, we assume that the transverse components of the nuclear spins (relative to the axis of the external magnetic field) 
evolve in a deterministic fashion, due only to their Larmor precession around the external field.  This assumption is justified, provided the total evolution time is not too long.  At longer times, one must take into account the changes in nuclear orientation due to the back action through the electron-nuclear coupling, as well as processes such as nuclear relaxation and diffusion, mediated by the nuclear dipole interactions.  In particular, the rotation of a given nucleus, mediated by the electron-nuclear coupling, depends not only on the hyperfine coupling strength of that nucleus, but also on the orientations of all other nuclei.  The hyperfine coupling, in turn, depends not only on the nuclear species, but also on the overlap with the electronic wave function, which varies with the position of the nucleus.

In order to obtain reliable results for longer times, it would be necessary to keep track separately of the evolutions of many composite spins, 
grouped according to the strengths of their hyperfine couplings, as well as their nuclear species, as described above.  Such calculations are possible\cite{Gullans2010,Gullans2012,Brataas2012},  but for long times they rapdily become computational expensive, and are beyond the scope of the present work.  Consequently, we have limited our calculations to time periods that are sufficiently short that the induced nuclear polarization is small compared to the random fluctuations in equilibrium. 

In experiments where the DNP protocol is repeated over millions of cycles, the nuclear polarization will eventually reach a steady state in which the polarization due to DNP is balanced by nuclear spin diffusion and relaxation processes arising from the nuclear dipole-dipole interaction or other mechanisms.   In this limit, there will be a new steady-state distribution
of nuclear spins, which may be quite different than the starting random distribution.  Although long-time distributions have been studied numerically in some limited cases [\onlinecite{Gullans2010,Gullans2012,Brataas2012}],  they have not been explored in the case of particular interest here, where commensurations between the nuclear Larmor periods and the cycle time are important. 
As far as the long-time steady state distribution of nuclear spin orientations remains unknown, our calculations, based on a  random nuclear distribution,  cannot be directly used to predict the outcome of a long-time  steady-state experiment.
Nevertheless, we believe that our results are applicable at least qualitatively to such experiments, and that our calculations give important insights into the origin of the commensurability effects observed in Ref [\onlinecite{Foletti2008}]. 

The rest of the paper is organized as follows. 
In section II we describe the physical system of interest: two electrons in a double quantum dot with hyperfine interaction to an ensemble of nuclear spins in the underlying crystal lattice. 
We describe the roles of hyperfine and spin-orbit coupling in the low energy subspace of the system, and outline the Hamiltonian governing the evolution of electron and nuclear spins.
In section III we describe the processes relevant for DNP in more detail. 
We present the semi-classical approximation, and discuss the
 roles of the Overhauser field, the spin-orbit interaction and
the transverse Knight field.  This semi-classical treatment is used in section IV,  where we present numerical results for specific manipulation 
protocols which yield an oscillatory dependence of DNP on cycle time, similar to the one observed by Foletti et al. 
In section V  we gain understanding of this behavior by analyzing various contributions to the DNP production rate. 
Finally, we present a discussion and summary of important results in section VI.

\section{Physical  Setup} 
In this section we briefly describe the physical system of interest and key terms in the Hamiltonian which governs its behavior.
This system has been described in great detail elsewhere\cite{Stepanenko2012}.
Here we summarize the most important features.

\subsection {Low energy subspace of a two-electron double quantum dot}

We consider two electrons in a gate-defined double quantum dot. 
Using the electrostatic gates which define the double-dot potential, the Coulomb-blockaded system can be tuned from a  ``(1,1)'' charge configuration, where electrostatics favor occupation of each dot by a single electron, to a ``(0,2)'' charge configuration where the right dot is doubly occupied in the ground state.
An in-plane magnetic field $\vec{B}=B\hat{z}$, which we assume has a negligible effect on the electrons' orbital motion, is applied along a direction which is defined to be the $z$-axis. 

Taking into account the Zeeman effect in the applied field, the two lowest-energy electronic spin states are mostly comprised of the triplet $\Ket{T_+} = \Ket{\uparrow_1\uparrow_2}$ and singlet $\Ket{S} = \frac{1}{\sqrt{2}}\left(\Ket{\uparrow_1\downarrow_2} - \Ket{\downarrow_1\uparrow_2}\right)$ configurations.
Here the arrows indicate the alignment of the electron spins with respect to the $z$ axis and the subscript indexes the electrons, 1 or 2.
In principle, the two-dimensional low energy subspace may also include admixture of the other two triplet spin states due to e.g. hyperfine or spin-orbit coupling.
However, here we assume that these interactions are weak compared with the electronic Zeeman energy, and therefore suppressed and negligible.
We study the regime where the electron temperature and all gate modulation frequencies are low enough that all other excited states are energetically inaccessible, and focus our attention on dynamics within the low-energy $S$-$T_+$  subspace.


Interdot tunnel coupling provides gate voltage control over the relative energies of the $\Ket{S}$ and $\Ket{T_+}$ states.
For a singlet spin configuration, tunnel coupling leads to hybridization of the $(1,1)$ and $(0,2)$ orbital states and opens an avoided crossing near the gate voltage where these two states are nominally degenerate.
Letting $\Delta$ stand for a parameter which controls the double dot detuning, or potential well asymmetry, we express the lower-energy hybridized singlet state as
\begin{equation}
\left|\Delta,S\right\rangle =a(\Delta)\left|(1,1)S\right\rangle +b(\Delta)\left|(0,2)S\right\rangle ,\label{eq:DeltaS}
\end{equation}
where $a$ and $b$ are smooth functions of the gate voltages which control the potential difference between the two dots.
In contrast, for the $\Ket{T_+}$ spin configuration, Pauli exclusion and a large single-well energy spacing only allow for a $(1,1)$ charge configuration.
Therefore, a large exchange energy difference between the singlet and triplet states can be achieved by tuning to the highly asymmetric potential regime where the (0,2) singlet state is heavily favored by electrostatics.
Alternatively, by tuning to a Coulomb-blockaded $(1,1)$ configuration, the triplet $\Ket{T_+}$ can be favored due to its lower Zeeman energy. 

\begin{figure}[t]
\includegraphics[ width=0.7\columnwidth]{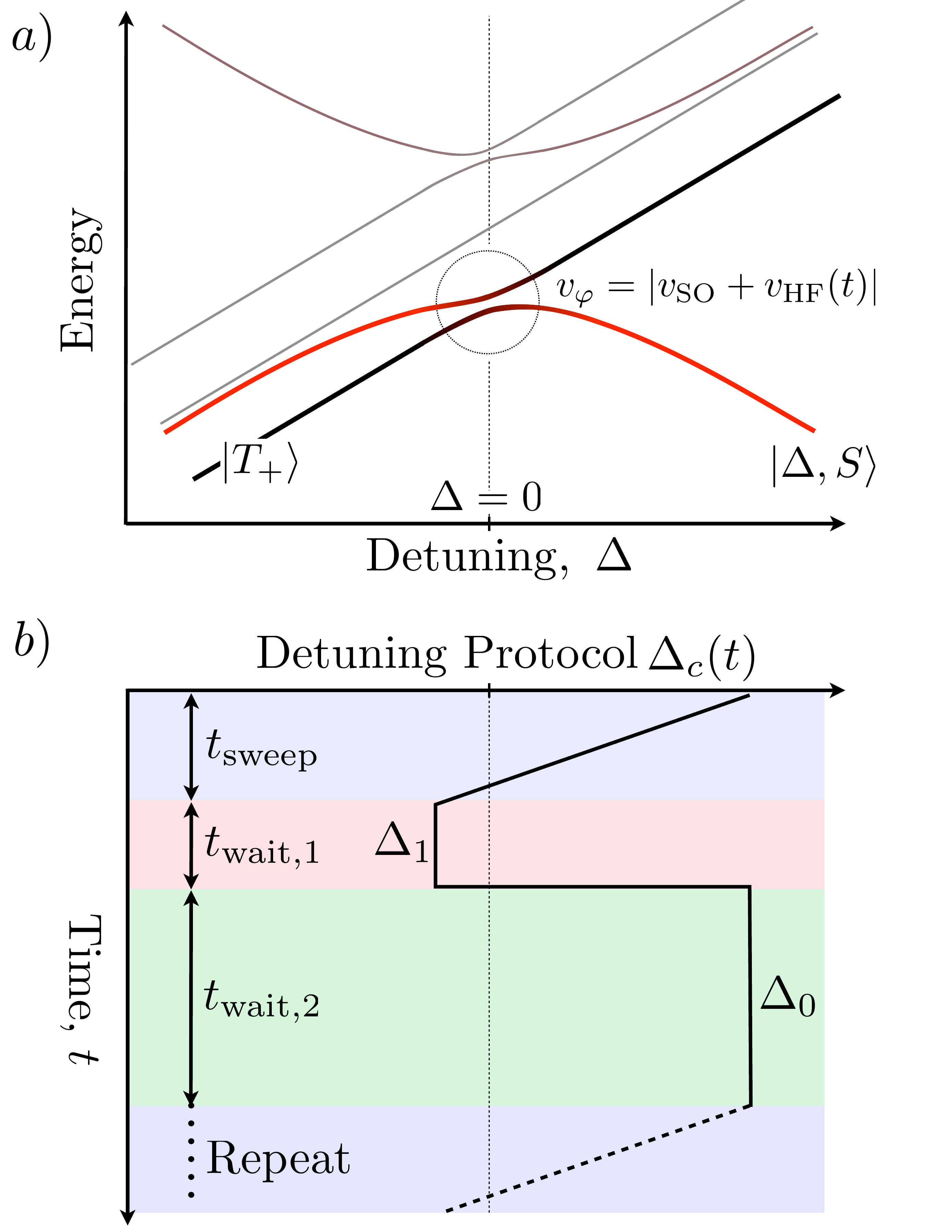}
\caption{Energy level diagram and detuning protocol. 
a) Two electron double quantum dot energy levels as a function of gate-voltage-controlled detuning, $\Delta_c$.
The detuning controls the potential well asymmetry, with the $(1,1)$ charge configuration 
favored for large negative values, and the $(0,2)$ charge configuration 
favored for large positive values.
Our work focuses on dynamics near the avoided crossing between the lower singlet level $\Ket{\Delta,S}$ and the triplet $\Ket{T_+}$, which arises due to an applied Zeeman field.
Near the singlet-triplet degeneracy point, defined to be $\Delta = 0$, the triplet and singlet states are mixed by hyperfine and spin-orbit interactions.
b) Time-dependent protocol for the controlled (i.e.~non-noisy) component of the detuning, $\Delta_c(t)$.
The details are described in Sec.~\ref{sec:Foletti}. 
}
\label{fig1}
\end{figure}

In this work we associate $\Delta$ with the energy difference between  $\Ket{T_+}$ and the lower energy singlet, $\Ket{\Delta, S}$, in the absence of hyperfine and spin-orbit interactions.
Indirectly, i.e.~through manipulation of gate voltages, $\Delta$ is the primary control parameter used in the DNP experiments that we discuss.
At a particular gate voltage configuration where exchange energy exactly compensates Zeeman energy, we have $\Delta = 0$.
Near this detuning, the eigenstates within the $S$-$T_+$ subspace exhibit an avoided level crossing due to the hyperfine and spin-orbit coupling matrix elements between $\Ket{\Delta,S}$ and $\Ket{T_+}$ (see Fig.\ref{fig1}).

Below we discuss the Hamiltonian which governs electron and nuclear spin dynamics near the $S$-$T_+$ avoided crossing. For simplicity of notation,  we shall generally denote $\Ket{\Delta,S}$ simply by $\Ket{S}$, except where we want to emphasize the dependence on $\Delta$. 
A detailed discussion of DNP and the semi-classical treatment of the nuclear spin dynamics is presented in the following sections. 

\subsection{
Effective Hamiltonian in the $S$-$T_+$ subspace}
Within the $S$-$T_+$ subspace, the system's evolution is governed by a Hamiltonian $\hat{H}  = H_\Delta + H_{\rm SO} + \hat{H}_{\rm HF} + \hat{H}_{\rm nuc}$.
Here $H_\Delta = \Delta \sigma^z$ describes the nominal detuning between the $\Ket{S}$ and $\Ket{T_+}$ states, in a basis where $\Ket{T_+}$ is listed first. 
The remaining terms, $H_{\rm SO}$ and $\hat{H}_{\rm HF}$ describe the effects of spin-orbit and hyperfine coupling, and $\hat{H}_{\rm nuc}$ contains the Hamiltonian terms of the nuclear spin system alone (hats are used to indicate operators which act on the nuclear spin system).

The spin-orbit Hamiltonian $H_{\rm SO}$ includes contributions from the 
Rashba and/or Dresselhaus terms \cite{Dresselhaus1955,Bychkov1983} for electrons in a GaAs two dimensional electron gas (2DEG), which
cause spin rotations to accompany orbital translations. 
In the Appendix we show that (see also Ref.~\onlinecite{Stepanenko2012})  
the spin-orbit interaction terms couple the states $\left|(1,1)T_{+}\right\rangle $ and $\left|(0,2)S\right\rangle $. 
This leads to a matrix element $v_{\rm SO} = \MatEl{T_+}{H_{\rm SO}}{\Delta, S}$ between the low-energy $\Ket{S}$ and $\Ket{T_+}$ states: 
\begin{equation}
v_{\rm SO}=b\left(\Delta\right)\frac{J}{\sqrt{2}}\frac{\left|\delta\vec{r}\right|}{\ell_{\rm SO}}\sin{\zeta}e^{i\varphi_{\rm SO}},\label{eq:vso}
\end{equation}
where $J$ is the tunnel coupling 
(which is half of the gap in the $\left|(1,1)S\right\rangle $ - $\left|(0,2)S\right\rangle $ anti-crossing), and $\left|\delta\mathbf{r}\right|=\left|\mathbf{r}_{L}-\mathbf{r}_{R}\right|$
is the distance between the centers of the two dots. 
Furthermore, $\ell_{\rm SO}$ is the spin-orbit length in the direction of the axis which connects the two dots, and $\zeta$ is the angle between   the direction of the magnetic field, $\vec{z}$, and the spin-orbit-induced spin-rotation axis for translations along this line (see appendix for precise definitions). 
The angle $\varphi_{\rm SO}$ depends on the choice of the axes $x$ and $y$. 
Viewing the two-dimensional $S$-$T_+$ subspace as that of an effective spin-1/2, the spin-orbit matrix element can be interpreted as a 
classical Zeeman field oriented in some direction in the $xy$-plane.
For the rest of the paper we orient the axes such that $v_{\rm SO}$ is real, i.e.~such that the effective spin-orbit field lies on the $x$ axis and $\varphi_{\rm SO}=0$.

In addition to spin-orbit coupling, the hyperfine interaction between the electron spins and nuclear spins of the host lattice also couples 
the $\Ket{S}$ and $\Ket{T_+}$ states.
The hyperfine coupling comes from the Fermi contact interaction, projected onto the states in the $S$-$T_+$ subspace\cite{Taylor2007}.
Because the interaction is {\it local}, it produces coupling only between $\Ket{T_+}$ and the $(1,1)$ orbital component of $\Ket{S}$.
Consequently, the hyperfine matrix element is proportional to $a(\Delta)$, see Eq.(\ref{eq:DeltaS}).
Thus the matrix elements of the hyperfine Hamiltonian $\hat{H}_{\rm HF}$ which couple $S$ and $T_+$ are related to those of the microscopic hyperfine Hamiltonian $\tilde{H}_{\rm HF}$ which acts purely within the $(1,1)$ subspace through the relation $\MatEl{T_+}{\hat{H}_{\rm HF}}{S} = a(\Delta)\MatEl{T_+}{\tilde{H}_{\rm HF}}{(1,1)S}$. Within the $(1,1)$ subspace, the hyperfine interaction is described by 
\begin{equation}
\tilde{H}_{\rm HF}=g^{*}\mu_{B}\sum_{d=L,R}\hat{\mathbf{B}}_{\rm nuc,d}\cdot\mathbf{S}_{\rm d},
\end{equation}
where $\hat{\mathbf{B}}_{\rm nuc,d}$ is the Overhauser-Zeeman field induced by the nuclei on the electron spin ${\mathbf S}_{\rm d}$ in dot d. 
Within each dot the Overhauser field receives contributions from all three nuclear species. 
We use the label $\alpha\in \{^{75}\rm As,\,^{69}\rm Ga,\,  ^{71}\rm Ga \}$ to denote the species, and write $\hat{\mathbf{B}}_{\rm nuc,d}=\sum_{\alpha}\hat{\mathbf{B}}_{\alpha,{\rm d}}$, with 
\begin{equation}\label{eq:B_alpha,d}
g^{*}\mu_{B}\hat{\mathbf{B}}_{\alpha,{\rm d}}\equiv\sum_{n_{\alpha}}A_{n_{\alpha},{\rm d}}\hat{\mathbf{I}}_{n_{\alpha}}.
\end{equation}
Here $\hat{\mathbf{I}}_{n_{\alpha}}$ is the spin operator for
the nucleus $n_{\alpha}$, with $n_{\alpha}$ running over all nuclei of species $\alpha$. 
The hyperfine coupling constant $A_{n_{\alpha},d}$ is equal 
to $\mathcal{A}_\alpha\left|\psi_{d}(\vec{r}{}_{n_{\alpha}})\right|^{2}$,
where $\left|\psi_{d}(\vec{r}_{n_{\alpha}})\right|^{2}$ is proportional to the electron density at the position $\vec{r}_{n_{\alpha}}$ of nucleus $n_\alpha$, and $\mathcal{A}_\alpha$ 
is the microscopic hyperfine coupling constant for nuclear spin species $\alpha$.
The values of the coupling constants, in the order  $\{^{75}\rm As,\,^{69}\rm Ga,\,  ^{71}\rm Ga \}$, are: $\mathcal{A}_{\alpha}/(g^*\mu_B) = \left(1.84\ \rm T,\,1.52\ \rm T,\,1.92\ \rm T\right) $. 
We assume identical density profiles $\left|\psi_{\rm d}(\vec{r})\right|^{2}$ for the states $\left|(1,1)T_{+}\right\rangle $ and
$\left|(1,1)S\right\rangle$. 

After projection back onto the $S$-$T_+$ low-energy subspace, the microscopic hyperfine coupling Hamiltonian $\tilde{H}_{\rm HF}$ produces the matrix element $\hat{v}_{\rm HF} = \MatEl{S}{\hat{H}_{\rm HF}}{T_+}$, given by\cite{Taylor2007,Erlingsson2005} 
\begin{equation}
\hat{v}_{\rm HF}=\frac{1}{2\sqrt{2}}a\left(\Delta\right)g^{*}\mu_{B}\sum_{\alpha}\Delta\hat{B}_{\alpha}^{+},\label{eq:v_HF}
\end{equation}
where $\Delta\hat{B}_{\alpha}^{+}=\hat{B}_{\alpha,L}^{+}-\hat{B}_{\alpha,R}^{+}$ and $\hat{B}_{{\rm \alpha},L(R)}^{+}=\hat{B}_{{\rm \alpha},L(R)}^{x}+i\hat{B}_{{\rm \alpha},L(R)}^{y}$.
The hat indicates that $\hat{v}_{\rm HF}$ is an operator which acts on the nuclear spin degrees of freedom.

For the free Hamiltonian of the nuclear spin system, $\hat{H}_{\rm nuc}$, we include only those
terms which account for nuclear Larmor precession 
in the applied field:
\begin{equation}
\hat{H}_{\rm nuc}=\sum_{\alpha=1}^{3}\omega_{\alpha}\hat{m}_{\alpha},\label{eq: H_nuc}
\end{equation}
where $\hat{m}_{\alpha}=\sum_{n_{\alpha}}\hat{I}_{n_{\alpha}}^{z}$ and $\omega_\alpha$ is the Larmor (angular) frequency of nuclear species $\alpha$.
Specifically, we ignore nuclear quadrupolar terms, which contribute to nuclear spin dephasing.
Below it will be useful to make the time-dependence associated with nuclear Larmor precession explicit.
We therefore switch to an interaction picture with respect to the Hamiltonian in Eq. (\ref{eq: H_nuc}), and write
\begin{equation}
\hat{B}_{\rm nuc,d}^{+}\left(t\right)=\sum_{\alpha=1}^{3}\hat{B}_{\alpha, {\rm d}}^{+}(0)e^{i\omega_{\alpha}t}.\label{eq:Larmor}
\end{equation}
Through Eq.~(\ref{eq:v_HF}), the hyperfine coupling matrix element $\hat{v}_{\rm HF}(t)$ between $\Ket{S}$ and $\Ket{T_+}$ states also acquires an explicit time-dependence in this interaction picture.

Collecting terms, we write the effective $S$-$T_+$ Hamiltonian in matrix form as 
\begin{equation}
\hat{H}=\left(\begin{array}{cc}
\frac{\Delta(t)}{2} & \hat{v}_{\rm HF}^{\dagger}(t)+v_{\rm SO}\\ \hat{v}_{\rm HF}(t)+v_{\rm SO}
 & -\frac{\Delta(t)}{2}
\end{array}\right).\label{eq: H}
\end{equation}
The Hamiltonian acts on a two-component spinor, $\psi\equiv\left[\psi_{T_{+}}(t),\ \psi_{S}(t)\right]^T$, which represents the electronic spin state, as well as the nuclear spin state. 
Together with Eq.(\ref{eq:Larmor}), this Hamiltonian defines the model that we will use to describe unitary evolution and DNP in this system.
%

Note that in Eq.~(\ref{eq: H}) we neglect the hyperfine coupling terms in the diagonal of the matrix, which describe the Overhauser shift of $\Ket{T_+}$ due to the $z$-component of nuclear polarization. 
In this work we focus on the DNP production rate for short times near the beginning of the pumping process, where the polarization and the induced Overhuser field are still small and can be neglected. 
These diagonal terms are also responsible for the Knight shift which modifies the nuclear Larmor frequencies depending on the electron spin state.
Fluctuations in this Knight field cause dephasing of nuclear precession and hence contribute to a loss of contrast in DNP oscillations. 
In neglecting these terms, we assume that other noise mechanisms dominate the dephasing of DNP oscillations (see below).

\section{The DNP production rate}\label{SchrodingerEquation}

Our goal in this work is to find the average change in nuclear polarization $\delta\langle\hat{m}_\alpha\rangle$ of each nuclear species, for a given detuning sweep protocol $\Delta(t)$. 
Depending on the specific experiment of interest, the function $\Delta(t)$ may describe one or many sweeps through the $S$-$T_+$ avoided crossing, 
with waiting periods at specific detunings between sweeps. 
In addition, for realistic experimental conditions $\Delta(t)$ generally contains a noisy part, due e.g.~to charge fluctuations, which must be taken into account. 
For the following discussion, we keep this function general. 
In section \ref{sec:Foletti} we discuss specific experimentally relevant protocols. 

As an operator equation, the rate of change of the polarization $\hat{m}_\alpha = \sum_{n_\alpha}\hat{I}^z_{n_\alpha}$ of species $\alpha$ is given by the Heisenberg equation of motion
\begin{equation}
\frac{d\hat{m}_\alpha}{dt} = -i[\hat{m}_\alpha, \hat{H}] = -i\left(\begin{array}{cc}0 & -\hat{v}_{\rm HF}^{(\alpha)\dagger}(t)\\ \hat{v}_{\rm HF}^{(\alpha)}(t) & 0 \end{array}\right)\label{EOM}.
\end{equation}
Here $\hat{v}_{\rm HF}^{(\alpha)}$ stands for the contribution to $\hat{v}_{\rm HF}$ in Eq.~(\ref{eq:v_HF}) coming from species $\alpha$ only.
Below we investigate 
Eq.~(\ref{EOM}) by means of a semi-classical approximation, which provides a clear physical picture for the resulting polarization dynamics.

\subsection{Semi-classical treatment of the Overhauser fields}
Due to the large number of nuclei $N_{\rm d}$ in each dot ($N_{\rm d} \sim 10^{6}$), it is a good approximation to replace the Overhauser fields $\hat{\mathbf{B}}_{\rm nuc,d}$ by classical vectors $\mathbf{B}_{\rm nuc,d}(t)$ 
which evolve smoothly in time. 
The validity of this approximation is based on 
the fact that the commutator $\left[\hat{B}_{\rm nuc,d}^{x},\hat{B}_{\rm nuc,d}^{y}\right]$ scales roughly as $|\hat{B}_{\rm nuc,d}^{z}|/N_{\rm d}$ 
and can be neglected. 
Making this approximation, we remove the hats in Eq.~(\ref{eq: H}) and replace the operators by their expectation values for a particular set of coherent nuclear spin states [keeping the explicit time dependence due to the Larmor precession, Eq.~(\ref{eq:Larmor})]:
\begin{eqnarray}
g^{*}\mu_{B}\hat{B}_{\alpha}^{+}\left(t\right) & \rightarrow & g^{*}\mu_{B}\left\langle \hat{B}_{\alpha}^{+}\left(t\right)\right\rangle \nonumber \\
 & = & v_{\alpha}e^{i\left(\theta_{0,\alpha}+\omega_{\alpha}t\right)},\label{eq: sc}
\end{eqnarray}
where $v_{\alpha}$ is real and positive and $\theta_{0,\alpha}$ is the initial azimuthal angle of the classical nuclear field at time $t=0$. 
The angle is measured relative to the $x$ axis of spin space, which is set by the direction of the effective spin-orbit field. 

 Because we are interested in the nuclear spin pumping rate close to equilibrium, where the nuclear spin state is initially completely random, we are free to choose how to average over the corresponding ensemble of nuclear spin states.
In making the semi-classical approximation, Eq.~(\ref{eq: sc}), we choose to first evaluate the dynamics for a particular pure, coherent state of the nuclear spins. 
We then average the resulting DNP over all such initial states, which corresponds to averaging over all possible magnitudes $\left\{v_\alpha\right\}$ and directions $\left\{\theta_{0,\alpha}\right\}$, to find the proper ensemble-averaged response.

Using Eq.~(\ref{eq: sc}), the hyperfine matrix element in Eq.~(\ref{eq: H}) is replaced by a sum of classical fields: 
\begin{equation}
\hat{v}_{\rm HF}(t) \rightarrow v_{\rm HF}(t) = \sum_{\alpha}v_{\alpha}e^{i\theta_{\alpha}\left(t\right)},\label{eq:v_HF_SC}
\end{equation}
where $\theta_{\alpha}(t)=\theta_{0,\alpha}+\omega_{\alpha}t$.   
In writing Eq.~(\ref{eq:v_HF_SC}) we used an approximation that $v_\alpha$ 
is independent of $m_\alpha$ and $t$. 
This approximation is valid as long as we focus on the evolution for small values of the nuclear polarization, 
and so long as the quantization axis for the nuclear spins is close to the $z$-axis, i.e.~the nuclear Zeeman field is much larger than the Knight field.

After replacing the Overhauser field operators by semi-classical fields, the electron spin evolution is governed by the Schr\"{o}dinger equation
\begin{equation}
i\frac{\partial\psi}{\partial t} = H_{\rm sc}\psi, \label{eq:shrodinger_sc}
\end{equation}
with
\begin{equation}
H_{\rm sc}(\{ \theta_{\alpha}\})=\left(\begin{array}{cc}
\frac{1}{2}\Delta(t) & v_{\rm HF}^*(t)+v_{\rm SO}\\
v_{\rm HF}(t)+v_{\rm SO} & -\frac{1}{2}\Delta(t)\end{array}\right).\label{eq:H_sc}
\end{equation}

Note that the matrix elements $v_{\rm SO}$ and $v_{\rm HF}$ in Eq.~(\ref{eq:H_sc}) depend on $\Delta$ through the coefficients $a$ and $b$ in Eqs.~(\ref{eq:v_HF}) and (\ref{eq:vso}).
Their values vary significantly when the detuning $\Delta$ changes on a scale comparable to the tunnel coupling $J$.
However, the physical processes described below are active near the $S$-$T_+$ avoided crossing (around zero detuning).
Because we work in the regime where $J$ is much bigger than the spin-orbit and hyperfine couplings, 
hereafter we use a simplified model in which $v_{\rm SO}$ and $v_{\rm HF}$ are taken to be constants, independent of $\Delta$ (cf.~Ref.~\onlinecite{Ribeiro2012}).

\subsection{DNP and the transverse Knight field 
\label{sub:perp Knight field}}

The semi-classical approximation immediately leads to a useful picture for understanding spin-dynamics in this system.
Writing the Hamiltonian in the form of Eq.~(\ref{eq:H_sc}) suggests an interpretation in which the two-component electronic wave function is viewed as that of an effective spin.
Indeed, this is a meaningful picture as the Pauli matrices $\sigma^x$ and $\sigma^y$ in this representation are directly proportional to the {\it differences} in the corresponding spin components: $\Delta S^x \equiv S^x_{L} - S^x_{R} \leftrightarrow \sqrt{2} \sigma^x$ and $\Delta S^y \equiv S^y_{L} - S^y_{R} \leftrightarrow \sqrt{2} \sigma^y$.
Similarly, the real and imaginary parts of the hyperfine matrix element $v_{\rm HF}$ in Eq.~(\ref{eq:H_sc}) are proportional to the Overhauser difference fields $\Delta B_{\rm nuc}^x$ and $\Delta B_{\rm nuc}^y$, and act as the corresponding $x$ and $y$ fields on the effective spin.
The back-action on the nuclei, which leads to DNP, then has a simple interpretation in terms of a torque exerted on the nuclear spins by the Knight field of the electrons.


A generic pure electron spin state in the $S$-$T_+$ subspace can be parametrized in terms of a unit vector $$\mathbf{n}=(\sin\eta\cos\xi,\sin\eta\sin\xi,\cos\eta),$$ where $\eta$ and $\xi$ are the polar and azimuthal angles of the effective spin on the Bloch sphere, respectively.
In two-component notation, the corresponding spinor is written as
$$\psi_{\mathbf{n}} = \left[\begin{array}{cc} \cos\frac{\eta}{2}e^{-i\frac{\xi}{2}}, & \sin\frac{\eta}{2}e^{i\frac{\xi}{2}}\end{array}\right]^{T}.$$
Given such a state for the electronic system, we now investigate how DNP is affected.

Replacing the nuclear spin operators in Eq.~(\ref{EOM}) by the corresponding classical fields, and taking an expectation value with respect to the electronic spin state represented by $\psi_{\mathbf{n}}$, we obtain a semi-classical equation of motion for the net (sum-field) nuclear polarization:
\begin{eqnarray}
\frac{dm_\alpha}{dt} & = & \psi^{\dagger}_{\mathbf{n}}
\left(\begin{array}{cc}
0 & iv_{\alpha}(t)e^{-i\theta_\alpha(t)}\\ 
-iv_{\alpha}^*(t)e^{i\theta_\alpha(t)} & 0 
\end{array}\right)\psi_{\mathbf{n}}\nonumber \\
\label{eq:dmdt_sc} & = & -\psi_{\mathbf{n}}^{\dagger}\frac{\partial H_{\rm sc}\left(\Delta(t),\left\{ \theta_{\alpha}\right\} \right)}{\partial\theta_{\alpha}}\psi_{\mathbf{n}}.
\end{eqnarray}
Equivalently, Eq.~(\ref{eq:dmdt_sc}) can be written in a convenient shorthand as 
\begin{equation}
\dot{m}_{\alpha}=v_{\alpha}\mathbf{z\cdot\left(\mathbf{n}\times\mathbf{b}_{\alpha} \right)},\label{eq: mdot}
\end{equation}
where $\mathbf{z}$ is a unit vector in the z direction, 
and $\mathbf{b}_\alpha$ is a unit vector parallel to the transverse component of the species-$\alpha$ Overhauser field, 
$$\mathbf{b_{\alpha}}=(\cos\theta_{\alpha},\sin\theta_{\alpha},0).$$

Equation (\ref{eq: mdot}) embodies the Bloch-equation-like nature of electron-nuclear spin dynamics.
The electron spins, described by the unit vector $\mathbf{n}$ produce a Knight field which acts on the nuclei.
Changes in the $z$-component of nuclear polarization arise from precession around the transverse part of this Knight field.

The key to understanding DNP in experiments is then to determine how the transverse components of electron spin evolve over specific gate sweep protocols.
In particular, it will be important to understand how the evolution of the Knight field itself is affected by the nuclear fields, which play a significant role in the 
Hamiltonian (\ref{eq:H_sc}) near the avoided crossing, $\Delta \approx 0$.
Back-action of the electrons onto the nuclei then provides a source of nonlinearity which can lead both to the build-up of large nuclear polarizations and to the transfer of polarization between different nuclear species.
Below we analyze these processes in more detail.

\subsection{Quantum and semi-classical expressions for the polarization change\label{sub:Quantum-and-sc-polarization}}





We restrict our discussion to the regime of small nuclear polarization, which is close to equilibrium, and assume that the DNP does not affect
the time dependent expectation value of the transverse field, $\left\langle \Delta\hat{B}_{\alpha}^{+}\left(t\right)\right\rangle =v_{\alpha}e^{i\left(\theta_{\alpha}+\omega_{\alpha}t\right)}$.
In this limit the equation of motion (\ref{eq:dmdt_sc}) for $m_\alpha$ and the equation of motion $\frac{d}{dt}\theta_\alpha(t)=\omega_{\alpha}$
for $\theta_\alpha$ result from the classical Hamiltonian $\psi^{\dagger}H_{\rm sc}\left(\left\{ \theta_{\alpha'}\right\} \right)\psi+H_{\rm nuc, sc}$,
with $\left\{m_\alpha,\theta_\alpha\right\} $ and $\left\{\psi, \left(i\psi^{\dagger}\right)\right\} $ the canonically-conjugate variables.



To obtain ensemble-averaged results, we must average
over many realizations of the initial conditions $\left\{v_{\alpha}e^{i\theta_{0,\alpha}}\right\}$.
These three complex vectors represent the 
the 
transverse Overhauser fields of the three nuclear species: $B^x_\alpha = v_\alpha \cos\theta_{0,\alpha}$ and $B^y_{\alpha} = v_\alpha \sin\theta_{0,\alpha}$.

Because the Overhauser field arises from a sum over many nuclear spins of a given species, 
we assume that all components are Gaussian-distributed in accordance with the central limit theorem. 
The distribution for the magnitude of $v_{\alpha}$ then satisfies 
\begin{equation}
\label{vdist}p(v_{\alpha}/\bar{v}_{\alpha})\frac{dv_{\alpha}}{\bar{v}_\alpha}=\frac{2v_{\alpha}}{\bar{v}_{\alpha}^{2}}e^{-v_{\alpha}^{2}/\bar{v}_{\alpha}^{2}}dv_{\alpha}, 
\end{equation}
with $\bar{v}_{\alpha}$ the rms value of the transverse Overhauser field for species $\alpha$. 
Note that the distribution $p(v_\alpha)$ is peaked around the value $\bar{v}_\alpha/\sqrt{2}$.
The initial phases $\theta_{0,\alpha}$ are uniformly distributed on the interval $-\pi$ to $\pi$. 

As we will discuss below, the interplay between nuclear spin coherence and electron spin dynamics is manifested in the high sensitivity of the nuclear polarization rate to the {\it phases} $\{\theta_{0,\alpha}\}$.
The phases control the interference of the Overhauser fields with each other and with the spin-orbit coupling, see Eqs.~(\ref{eq:v_HF_SC}) and (\ref{eq:H_sc}).
The dependence on the Overhauser field magnitudes is much weaker. 
Therefore below we focus on the {\it phase-averaged} nuclear polarization rate for fixed, typical values $\{v_\alpha\}$.
In Sec.~\ref{Discussion} we discuss the effects of averaging over $\left\{ v_{\alpha}\right\}$, and explain why we do not expect it to significantly change our results.

For a given experimental cycle $i$ running from time $t_i$ to $t_{i+1}=t_i + t_{\rm cyc}$, where $t_{\rm cyc}$ is the cycle time, the {\it phase-averaged} change in nuclear polarization for
a given realization of transverse Overhauser field magnitudes $\left\{v_{\alpha}\right\}$ is given by 
\begin{equation}
\label{eq:m_bar_v} \overline{\delta m}_\alpha(\{ v_{\alpha'}; i\}) = \oint \frac{d^3\theta_0}{(2\pi)^3}\, \int_{t_{i}}^{t_{i+1}} \frac{dm_\alpha}{dt}(\{ v_{\alpha'}\}, \{\theta_{0,\alpha'}\})dt,
\end{equation}
where $d^3\theta_0 = \prod_{\alpha'}d\theta_{0,\alpha'}$.
As is evident in Eq.~(\ref{eq:dmdt_sc}), the polarization rate $\frac{dm_\alpha}{dt}$ depends on the electronic spinor $\psi$.
In Eq.~(\ref{eq:m_bar_v}) we assume that the electronic state is initialized to $\Ket{S}$ at time $t_0$, independent of the nuclear state.
Therefore the change in nuclear polarization obtained in cycle $i$ implicitly depends on the electronic evolution for all times from $t_0$ up to $t_i + t_{\rm cyc}$.

Below we show that, under realistic conditions in the presence of relaxation and decoherence, the electronic state loses memory of its initial state after several sweeps.
After this point, the phase-averaged change in nuclear polarization per sweep, Eq.~(\ref{eq:m_bar_v}), loses its dependence on the cycle index $i$.
Our aim is to describe how the resulting steady-state polarization rates depend on the parameters of experimentally-relevant qubit manipulation protocols.


\section{Polarization under experimental sweep protocols\label{sec:Foletti}}

In this section we study how dynamical nuclear polarization develops under specific experimental protocols.
We focus on cases which feature long waiting periods near the $S$-$T_+$ anti-crossing.
In order to capture important features of the experiments, we include the effect of charge noise in the gates or in the nearby 2DEG. 
This noise generally causes the detuning $\Delta$ to fluctuate randomly in time.
This produces electron spin dephasing and relaxation on top of the unitary dynamics expected for the case of a deterministic time-dependent detuning profile.


In order to investigate the build-up of nuclear polarization via Eqs.~(\ref{eq:dmdt_sc}) and (\ref{eq:m_bar_v}), we must solve for the electron spin evolution in the presence of a classical time-dependent field produced by the nuclear spins.
On top of this, we 
include the effects of random noise in the detuning, which is responsible for electronic singlet-triplet dephasing.
To this end, we split the time-dependent detuning into deterministic and noisy parts: $\Delta(t) = \Delta_{c}(t)+\Delta_{n}(t)$. 

The experimentally-controlled component of the detuning, $\Delta_{c}(t)$, is a prescribed, deterministic function, which is periodic in time for $t>0$.
Hereafter we refer to $\Delta_c(t)$ as the ``DNP protocol.'' 
Each periodic repetition in $\Delta_c$ is one ``cycle'' of the protocol. 
The cycles are generally comprised of ``sweeps,'' during which $\Delta_c$ rapidly changes, taking the system through the $S$-$T_+$ anti-crossing, and ``waiting periods,'' in which $\Delta_c$ is held constant for a prescribed interval between sweeps. 

We focus in particular on a specific class of DNP protocols similar to those used in Ref.~\onlinecite{Foletti2008}, and depicted in Fig.~\ref{fig1}b.
At time $t=0$ the electronic spin system is initialized in the singlet state at large positive detuning, $\Delta_{c}(t = 0) = \Delta_{0}$, with $\Delta_0 \gg v_{\rm max}$, where $v_{\rm max} = v_{\rm SO} + \sum_{\alpha}v_{\alpha}$. 
The detuning is then rapidly swept over a short time $t_{\rm sweep}$ 
to a value $\Delta_c(t = t_{\rm sweep}) = \Delta_{1}$, bringing the system close to the $S$-$T_{+}$ anti-crossing point.
Here the detuning is held fixed for a waiting time $t_{\rm wait,1}$: $\Delta_c(t) = \Delta_1$ for $t_{\rm sweep} \le t \le t_{\rm sweep} + t_{\rm wait,1}$. 
Then, the detuning is ramped very quickly back to the initial, large positive detuning $\Delta_0$. 
For simplicity we assume that the return sweep after the waiting time $t_{\rm wait, 1}$ is {\it instantaneous}.
Finally, the detuning is held fixed again for a second (typically much longer) waiting period of duration $t_{\rm wait, 2}$: $\Delta_c(t) = \Delta_0$ for $t_{\rm sweep} + t_{\rm wait,1} \le t \le t_{\rm cycle}$, where 
$t_{\rm{cycle}}=t_{\rm{sweep}}+t_{\rm wait,1}+t_{\rm wait,2}$ is the cycle time.
This cycle is then repeated over and over again.


The function $\Delta_{n}(t)$ is taken to be random white noise, with an amplitude much smaller than the typical range of values covered by the sweeps of $\Delta_c(t)$. 
Note that, in principle, even the case of a noisy detuning function is covered by the discussion above in Sec.~\ref{SchrodingerEquation}. 
This is true because, for any particular realization of the detuning function $\Delta(t)$, the semi-classical Schr\"{o}dinger equation (\ref{eq:shrodinger_sc}) produces unitary evolution of the two-electron spin state. 
That is, for each particular realization of the classical noise $\Delta_n(t)$, the electronic state remains pure.
However, we are interested in the polarization rate resulting from an average over many cycles.
Because the noise realization $\Delta_n(t)$ is different in every cycle, we must average the results with respect to the distribution from which $\Delta_n$ is drawn. 
As described in the following subsection, this averaging is conveniently treated in a framework where the electronic spin state is represented by a $2\times 2$ density matrix.

\subsection{Non-unitary electron spin evolution due to noise \label{sub:Non Hermitian part}}
In the absence of noise, the electron spins remain in a pure state represented by the two-component spinor $\psi(t)$, with corresponding density matrix $\rho(t)= \psi\psi^{\dagger}$. 
What happens in the presence of noise?
Consider first a waiting period in which $\Delta_c$ is held constant at a value where 
there is a large energy difference between the states $\Ket{T_+}$ 
and $\Ket{\Delta,S}$.
Due to the different charge distributions in the singlet and triplet states, noise-induced tilts of the double-well quantum dot potential cause the relative energies of these states to fluctuate. 
Thus detuning noise causes dephasing of superpositions of the singlet and triplet states, which is characterized by decay of the off-diagonal matrix elements of $\rho$ in the $T_+$-$S$ basis.
For small values of $\Delta_c$, the eigenvectors of the electronic semiclassical Hamiltonian $H_{\rm sc}$ are superpositions of the states $\left|T_{+}\right\rangle$ and $\left|\Delta,S\right\rangle$.
Here the same detuning noise acts via an off-diagonal operator in the energy-basis, and thus causes both dephasing and incoherent transitions between eigenstates.

We model this behavior phenomenologically via a Lindblad master equation for the electronic density matrix, 
\begin{equation}
\label{eq:lindblad_eqution}\dot{\rho} = i\left[\rho,H_{\rm sc}(\Delta_{c}(t),\left\{ \theta_{\alpha}\right\})\right] + L\rho L^{\dagger}-\frac{1}{2}\left\{ L^{\dagger}L,\rho\right\} ,
\end{equation}
where $L$ is the Lindblad jump operator
\begin{equation}
L = \frac{\gamma}{2}
\left(\begin{array}{cc}
1 & 0\\
0 & -1\end{array}\right),\label{eq:dephasing_lindblad}
\end{equation}
and $\gamma$ is a phenomenological decoherence rate which is related to the noise intensity of $\Delta_n(t)$.
Note that $L$ in Eq.~(\ref{eq:dephasing_lindblad}) is given in the singlet-triplet basis. 
As described above, the corresponding jump terms cause pure dephasing whenever the eigenvectors of the Hamiltonian $H_{\rm sc}$ are approximately given by $\Ket{T_+}$ and $\Ket{\Delta,S}$. 
However, near the $S$-$T_+$ anti-crossing, the eigenvectors of $H_{\rm sc}$ are superpositions of $S$ and $T_+$ and $L$ induces transitions between them. 

 Below we consider DNP protocols where the sweeps occur on times much shorter than $\gamma^{-1}$, while the waiting times are typically much longer than $\gamma^{-1}$.
In this case, assuming that $\gamma$ is much smaller than the energy splitting of $H_{\rm sc}$ at the waiting point, the density matrix approaches a diagonal form (in the instantaneous eigenbasis of $H_{\rm sc}$) during the waiting periods. 
Denote the instantaneous eigenvectors of $H_{\rm sc}$ at the waiting point $\Delta$ by $\chi_{\pm}(\Delta,\{\theta_\alpha\}) = \left(\chi_{S,\pm},\chi_{T_{+},\pm}\right)^{T}$.
At the end of the waiting period, the density matrix is approximately given by
\begin{equation*}
\rho = P_{+}\chi_{+}\chi_{+}^{\dagger} + P_{-}\chi_{-}\chi_{-}^{\dagger},
\end{equation*}
where $P_+$ and $P_-$ are the occupation probabilities of the states represented by $\chi_+$ and $\chi_-$.

In many experiments, the DNP protocol involves a long waiting period $t_{\rm{wait},2}$ at large detuning at the end of each cycle.
After such a waiting period, the diagonal entries of the density matrix correspond to the occupation probabilities of the singlet and triplet states.
These probabilities depend on the initial density matrix at the beginning of a given cycle, and on the detailed evolution during the cycle.
The latter crucially depends on the Overhauser fields 
via their influence on the splitting of the $S$-$T_{+}$ anti-crossing. 
As noted earlier, we assume that the Overhauser field magnitudes $\{v_\alpha\}$ remain constant and that the corresponding phases evolve according to simple Larmor precession, $\theta_\alpha(t)=\theta_{\alpha,0}+\omega_\alpha t$.
Therefore we find that the electronic density matrix at the end of the $i$-th cycle depends on the evolution through the flip probability $F_i = F(\{\theta_{\alpha}(t_{i})\})$,
where $\{ \theta_{\alpha}(t_{i})\}$ are the phases of the Overhauser fields at the beginning of the cycle. 

In principle, the rate $\gamma$ appearing in Eq.~(\ref{eq:lindblad_eqution}) depends on the detuning $\Delta$ via the detuning-dependent susceptibility of the singlet state $\Ket{\Delta, S}$.  
Similar to the spin-orbit and hyperfine matrix elements, we take $\gamma$ to be independent of $\Delta$ throughout this work.

Another important ingredient for understanding experiments is the relaxation from triplet to singlet that occurs while the system is held at large detuning. 
Here, relaxation typically occurs via phonon-mediated hyperfine or spin-orbit spin-flip processes, or via virtual processes which mediate exchange with the leads.
In the absence of such relaxation, the electronic state would approach an incoherent 50/50 mixture of singlet and triplet in the steady-state.
Owing to the symmetry between the singlet-to-triplet and triplet-to-singlet spin flip probabilities, the nuclear polarization rate vanishes for the 50/50 totally mixed electronic state.
Relaxation maintains an imbalance between the singlet and triplet occupation probabilities, thus allowing a continuous pumping of angular momentum into the nuclear subsystem.

In this work we model singlet-triplet relaxation by introducing a decay process that takes the qubit state from $\Ket{T_+}$ to $\Ket{S}$ with probability $\Lambda$ at the end of each cycle.
Here we assume that the cycle finishes with a waiting period at large positive detuning (where the ground state is approximately $\Ket{(0,2)S}$). Experimental protocols in which the electronic state is intentionally reloaded to $\Ket{S}$ after each cycle are described by $\Lambda = 1$. 
 
Here, a question arises regarding a possibile change of nuclear polarization accompanying the relaxation events from $\Ket{T_+}$ to $\Ket{S}$ at large positive $\Delta$. 
Such a polarization change is expected if spin relaxation is mediated by hyperfine coupling accompanied by phonon emission. 
In this case, the direction of the polarization change is opposite to that produced by a single gate sweep through the $\Ket{S}$-$\Ket{T_+}$ crossing, with the system initially in the singlet state.
On the other hand, if the relaxation occurs either due to spin exchange with the leads, or due to spin-orbit coupling assisted phonon emission, then no polarization occurs. 
In the general case, when both spin-orbit and hyperfine interactions are present,
the precise amount of polarization transfer per relaxation event depends on the the value of the detuning at the waiting point, on the magnetic field, and may also depend on the instantaneous orientations of the hyperfine fields.  However, a full investigation of these processes is beyond the scope of this work.
Here we assume that all relaxation processes occur through nuclear-spin-independent mechanisms, i.e.~that nuclear polarization is preserved during electron spin relaxation.

Taking into account the combination of strong dephasing and relaxation during the waiting time at the end of each cycle, the electronic state loses its memory of its initial state after many cycles.  
After a large number of cycles $i$, the density matrix at the beginning of a given cycle $i + 1$ is nearly diagonal, with singlet and triplet occupation probabilities $P_
S$ and $P_{T_+}$.
Here $P_S$ and $P_{T_+}$  are functions that depend on the values of the Overhauser field phases $\{\theta_{\alpha}(t_{i+1})\}$, evaluated at the beginning of the cycle. 
From cycle to cycle, the occupation probabilities (taken at the start of the cycle) evolve according to the discrete map
\begin{widetext}
\begin{equation}
\left.\left(\begin{array}{c}
P_{T_+}\\
P_{S}\end{array}\right)\right|_{\left\{ \theta_{0,\alpha}+\omega_{\alpha}t_{i+1}\right\} }=\left(\begin{array}{cc}
1-\Lambda & 0\\
\Lambda & 1\end{array}\right)\cdot\left(\begin{array}{cc}
1-F_{i} & F_{i}\\
F_{i} & 1-F_{i}\end{array}\right)\cdot\left.\left(\begin{array}{c}
P_{T_+}\\
P_{S}\end{array}\right)\right|_{\left\{ \theta_{0,\alpha}+\omega_{\alpha}t_{i}\right\} }.\label{eq: Gamma*F_i}
\end{equation}
\end{widetext}

For $\Lambda = 0$, we find a steady state with $P_{S} = P_{T_+}=\frac{1}{2}$. 
Indeed, a nonzero decay probability $\Lambda$ is necessary for maintaining a nontrivial electronic steady state, with 
$P_{S}\neq P_{T_+}$.
As we now show, this asymmetry, together with the dependence of the flip probability $F$ on the Overhauser field phases $\{\theta_{\alpha}(t_{i})\}$,
gives rise to a steady state production of DNP.
The DNP production rate is sensitive to commensurations between nuclear Larmor precession and the gate sweeps, and survives the averaging over all initial phases $\{ \theta_{0,\alpha}\}$.

\subsection{Oscillatory dependence of DNP on cycle time\label{sec:Folleti's osc. simulation}}

We now demonstrate that our model, applied to the DNP protocol in Fig.~\ref{fig1}, produces an oscillatory response of the DNP production rate on cycle time which mimics the behavior observed the experiments of Ref.~\onlinecite{Foletti2008}.
We mostly focus on the case $\Delta_1 < 0 $, where each sweep takes the system through the avoided crossing.
Furthermore, we take $\frac{d\Delta_c}{dt} \gtrsim v_{\rm max}^2$. 
This latter condition means that the singlet-triplet transition probability for a single sweep is small, but not insignificant. 
Below we first present numerical results from this model, followed by an in-depth analytical treatment.

\begin{figure}[t]
\includegraphics[width=3.0 in]{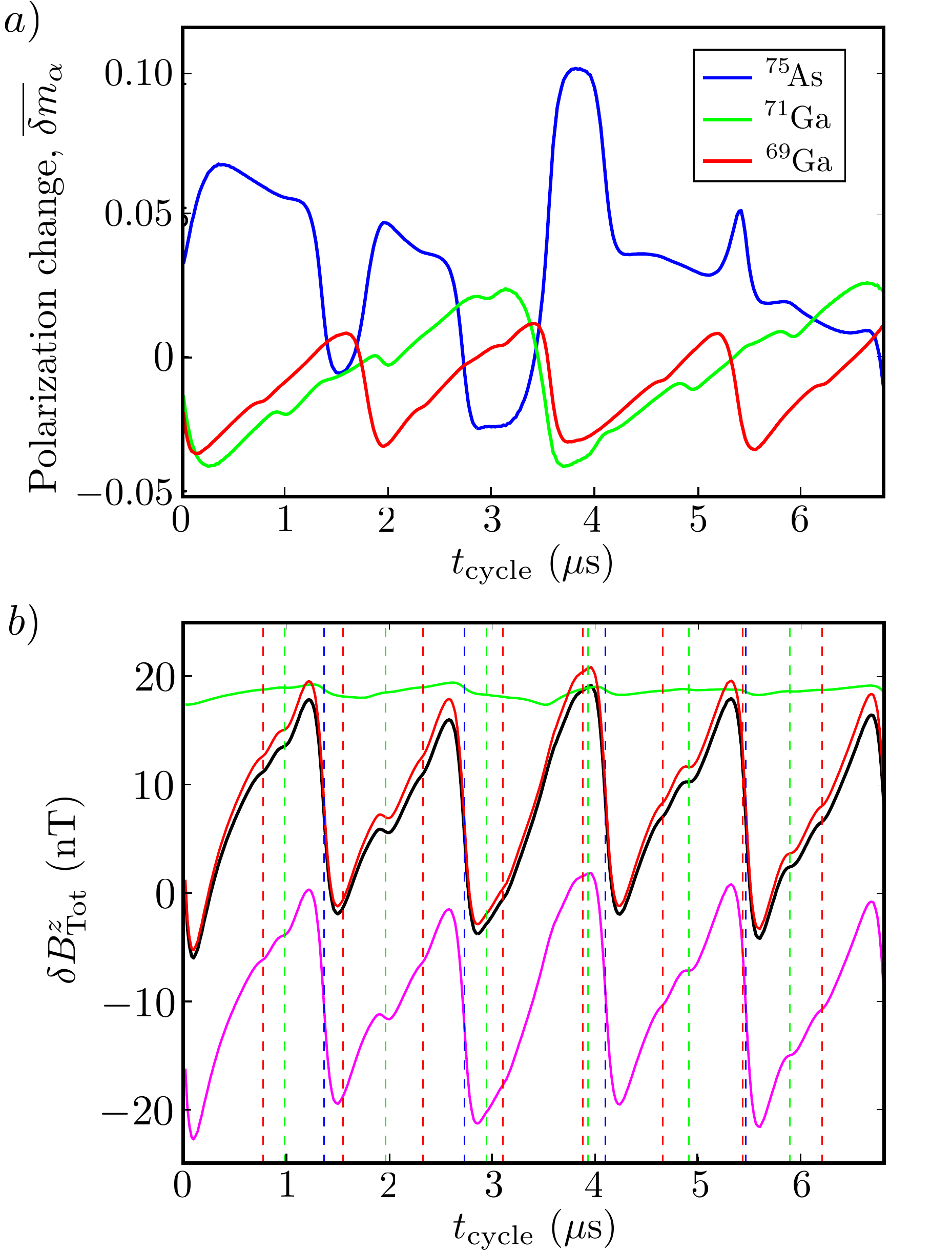}
\caption{
Phase-averaged steady state polarization per cycle, Eq.~(\ref{eq:m_bar_v}), for the experimental protocol discussed in the text with parameter values given below. 
a) $\overline{\delta m}_\alpha$ {\it vs.} total cycle time $t_{\rm cycle}$ (with $t_{\rm sweep}$ and $t_{\rm wait,1}$ fixed) for each of the three species. 
b) Change in the total Overhauser field per cycle, $\delta B_{\rm Tot}^z = \sum_{\alpha,d}\delta B^z_{\alpha,d}\approx \frac1{N_d}\sum_{\alpha}\mathcal{A}_\alpha\overline{\delta m}_\alpha$. 
Here we use a model with uniform electron density, which ignores spatial variations of $\left|\psi_{d}(\vec{r}_{n_{\alpha}})\right|^{2}$ in Eq.~(\ref{eq:B_alpha,d}).
The black line shows the result from the full simulation. 
The green and purple lines show the geometrical and (incoherent) dynamical contributions 
(see section \ref{sec:DNP}), while the red line shows their sum.
The three vertical dashed lines mark the times that are commensurate with the Larmor periods of each of the species, color-coded according to the legend in the top panel.
Parameter values: $\gamma^{-1}=100$ ns, $\Delta_0=40v_{\rm max}$, $\Delta_{1} = -3v_{\rm max}$, where $v_{\rm max} = v_{\rm SO} + \sum_{\alpha}v_{\alpha}$, $\frac{d}{dt}\Delta=-40v_{\rm max}^2$,  $t_{\rm wait, 1}=400$ ns, and $\Lambda = 0.1$.
The nuclear Larmor (angular) frequencies were  $\omega_\alpha=(4.6,6.4,8.1)\,\times 10^6$ rad/s, corresponding to $B_{\rm ext} = 0.1$ T, and the transverse Overhauser field amplitudes were taken to be  $v_\alpha=(4.56,2.26,1.90)\, \times 10^7$ rad/s, corresponding to typical values for $N_d \sim 4 \times 10^6$ nuclei. The spin-orbit matrix element is taken to be $v_{\rm{SO}}=\frac14\sum_\alpha v_\alpha=2.18\times10^7$ $\rm rad/s$.}
\label{fig:DNP}
\end{figure}
In Fig.~\ref{fig:DNP} we plot the steady state phase-averaged polarization change per cycle, for the protocol described above.
The results were obtained by numerically integrating Eq.~(\ref{eq:m_bar_v}), with $\frac{dm_\alpha}{dt}$ calculated from the electron spin density matrix $\rho$ via 
\begin{equation}
 \frac{dm_{\alpha}}{dt} = -{\rm Tr}\left[\rho\frac{\partial H_{\rm sc}}{\partial\theta_{\alpha}}\right],\label{eq:dmdt_rho_v1}
 \end{equation}
where $H_{\rm sc} = H_{\rm sc}\left(\Delta_{c}(t),\{\theta_{\alpha}\}\right)$, see Eq.~(\ref{eq:H_sc}).
The evolution of $\rho$ over the cycle is calculated according to Eq.~(\ref{eq:lindblad_eqution}), with an initial state equal to the steady state solution of Eq.~(\ref{eq: Gamma*F_i}). 
The parameters are listed in the caption. 
Rapid drops in the net Overhauser shift are clearly visible (Fig.~\ref{fig:DNP}b) whenever the total cycle time passes through an integer multiple of the Larmor period for $^{75}$As.  Smaller drops are also visible at multiples of the Larmor periods for $^{71}$Ga and $^{69}$Ga.
This behavior bears a strong resemblance to the 
oscillations found experimentally in Ref.~\onlinecite{Foletti2008}.
 Note that the quasi-periodic structure is present {\it after} averaging over all initial phases 
$\left\{ \theta_{0,\alpha}\right\}$ of the Overhauser fields.

\begin{figure}[h]
\includegraphics[width=3.0 in]{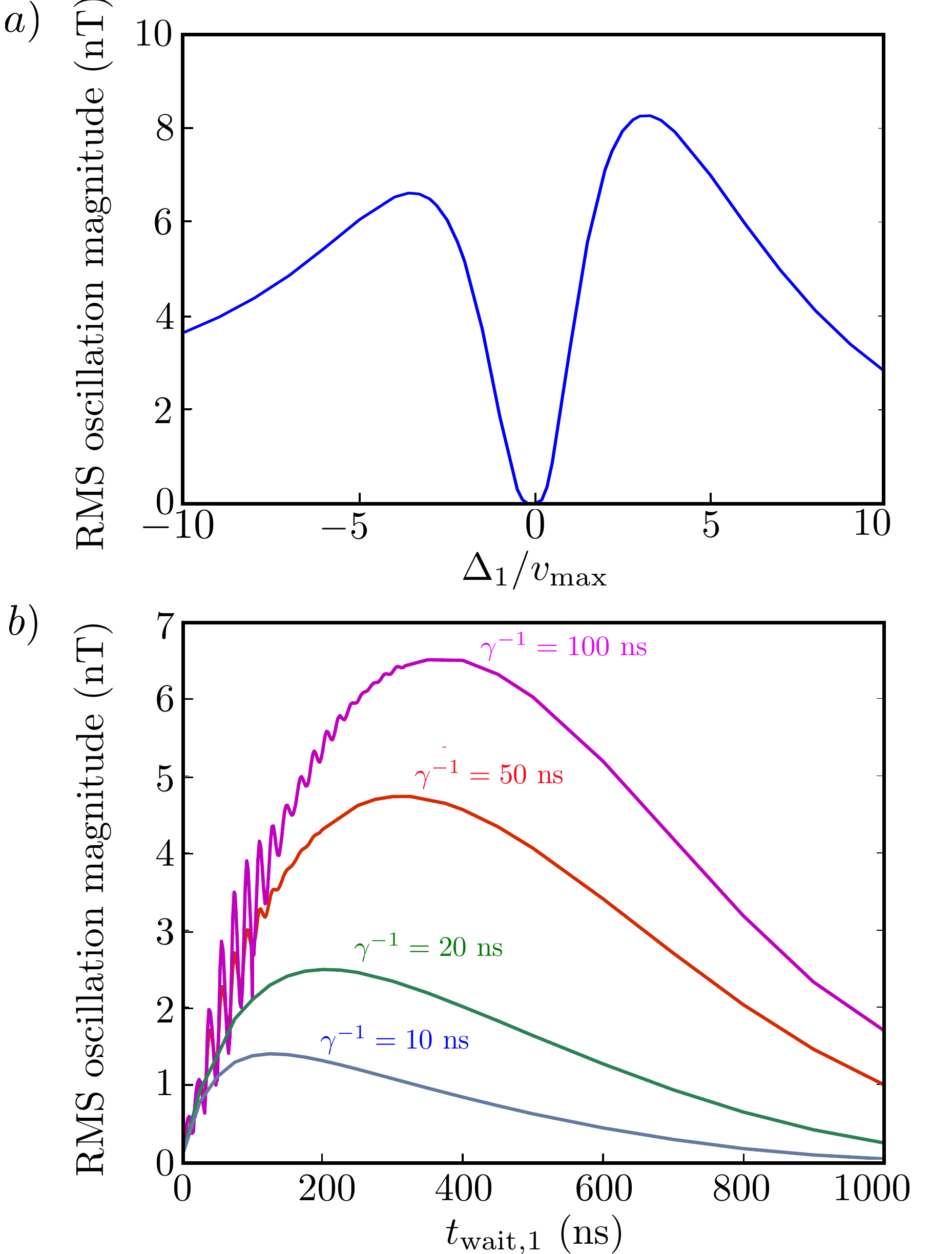}
\caption{Dependence of the oscillation amplitude on the waiting time position and duration. 
For each set of parameters, we extract the root-mean-squared (rms) magnitude of the oscillations in 
the trace $\delta B^z_{\rm Tot}(t_{\rm cycle})$, see e.g.~black line in Fig.~\ref{fig:DNP}b, calculated relative to its mean value. 
In all cases, we run the simulation for a range of $t_{\rm cycle}$ values 
which exactly spans 7 periods of the 
the $^{75}$As Larmor precession.
(top) Oscillation magnitude {\it vs.}~position of the waiting point $\Delta_{1}$. 
(bottom) Oscillation magnitude {\it vs.}~the waiting time near the crossing, $t_{\rm wait,1}$.
Other parameters are as given in the caption of Fig.~\ref{fig:DNP}.} 
\label{fig:osc_vs_t1}
\end{figure}
These intriguing oscillations have several peculiar properties.
First, note that in the species-resolved DNP curves in Fig.~\ref{fig:DNP}a the largest amplitude signal comes from $^{75}$As, but it does not show any distinct periodic features. 
Nonetheless, the net Overhauser field which is induced by the {\it total} DNP of all the species, plotted in Fig.~\ref{fig:DNP}b, shows pronounced, approximately periodic features at times commensurate with the $^{75}$As Larmor period. 

In addition, the oscillation amplitude displays an interesting dependence on the parameters of the protocol.
Figure \ref{fig:osc_vs_t1}a shows how this amplitude depends on the detuning $\Delta_1$ during the waiting time $t_{\rm wait,1}$ near the $S$-$T_+$ crossing. 
Note that the oscillations appear for  $\Delta_1 \neq 0$, i.e.~the system must be held slightly away from the crossing in order for the oscillations to show-up.
Figure~\ref{fig:osc_vs_t1}b shows the dependence of the oscillation amplitude on the waiting period $t_{\rm wait,1}$, for three values of dephasing rate $\gamma$. 
For short waiting times, the oscillation intensity varies rapidly with waiting time due to electron spin coherence.
The time scale associated with these variations corresponds to the inverse of the singlet-triplet splitting.
Interestingly, the oscillation intensity is maximal for longer waiting times, where $\gamma t_{\rm wait,1}\sim 4-10$. 
This indicates that the main contribution to the DNP oscillations comes from times long after electron spin coherence is lost.

\begin{figure}[t]
\includegraphics[width=3.0 in]{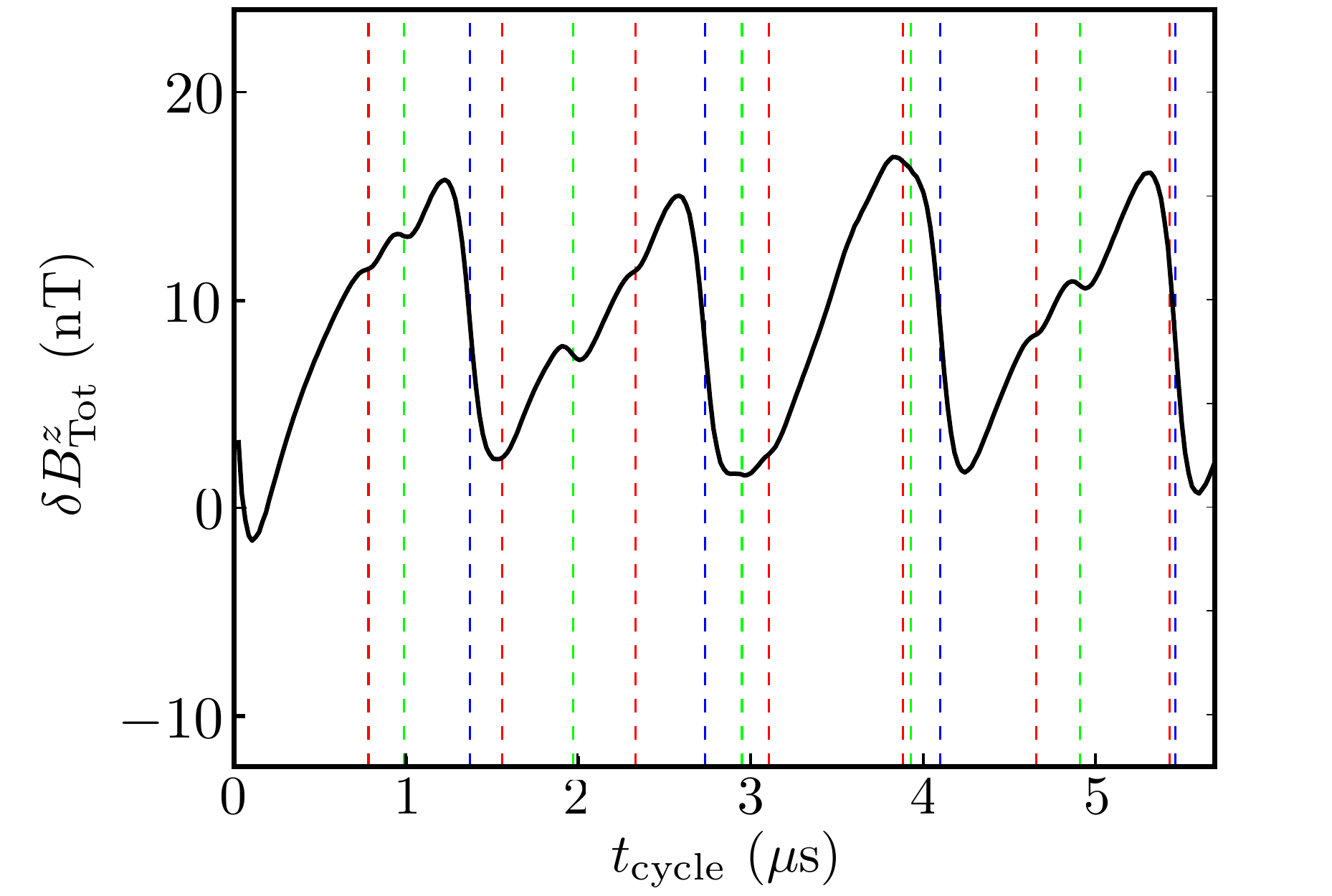}
\caption{
Robustness of oscillations to averaging of Overhauser field magnitudes. 
Phase-averaged DNP, Eq.~(\ref{eq:m_bar_v}), averaged also with respect to the magnitudes of the three Overhauser fields.
We average using a discrete distribution of 5 values for each species (125 values in total), which mimics the distribution in Eq.~(\ref{vdist}) (see text). 
  All other simulation parameters are as in Fig.~\ref{fig:DNP}. The oscillations closely resemble those in Fig.~\ref{fig:DNP}, 
 with an amplitude diminished by approximately 30\%.
}
\label{fig:DNP_bar}
\end{figure}

Note that the oscillations shown in Fig.~\ref{fig:DNP} were obtained for a fixed set of Overhauser field magnitudes $\{v_{\alpha}\}$. 
In Fig.~\ref{fig:DNP_bar} we demonstrate that the oscillatory behavior survives averaging over the three Overhauser field magnitudes.
Because the averaging according to Eq.~(\ref{vdist}) is computationally quite expensive, we implement the averaging with a discrete distribution, taking 5 values for each of the 3 Overhauser field magnitudes (125 total choices).
The values are chosen to mimic the distribution $p(v_\alpha/\bar{v}_\alpha)$ in Eq.~(\ref{vdist}), when sampled with equal probabilities $p_i=\frac15$. The five possible magnitudes of each Overhauser field, $v_{\alpha,i}$, were determined by preserving the expectation value, via the condition $\frac15v_{\alpha,i} =\bar{v}_\alpha\int_{x_{i-1}}^{x_i}{xp(x)dx}$, where the integration bounds satisfy $x_0=0$ and $\int_{x_{i-1}}^{x_i}{p(x)dx}=\frac15$ for $i=1,\ldots,5$.

In the theoretical analysis below we focus first on the case of fixed Overhauser field magnitudes, and explain in detail how the oscillations arise after averaging over all orientations $\{\theta_{0,\alpha}\}$.
At the end, we will 
discuss why the oscillations 
survive the averaging over the Overhauser field magnitudes as well.


\section{Analysis of the DNP oscillations\label{sec:DNP}}

In order to elucidate the physics underlying the oscillatory dependence of DNP on the cycle time, we now analyze various contributions to the polarization in more detail.
First, we transform to a rotating frame in which the phase of the off-diagonal matrix element in the semi-classical electron spin Hamiltonian, Eq.~(\ref{eq:H_sc}), is constant in time.
We then identify a ``geometrical'' contribution  to the  DNP which arises from the transformation to the rotating frame, and a ``dynamical'' contribution which is generated by the Hamiltonian in this rotating frame.

Before performing the transformation to the rotating frame, we highlight a remarkable fact about the phase-averaged polarization change $\overline{\delta m}_\alpha$, defined in Eq.~(\ref{eq:m_bar_v}).
Reversing the order of integration over time and phase variables in Eq.~(\ref{eq:m_bar_v}), we define the phase-averaged polarization {\it rate}
\begin{equation}
\label{dmdt_phaseavg} \overline{\frac{dm_\alpha}{dt}}(\{ v_{\alpha}\}) = \frac{1}{(2\pi)^3}\oint d^3\theta_0\frac{dm_{\alpha}}{dt}(\{ v_{\alpha'}\}, \{\theta_{0,\alpha'}\}).
\end{equation}
For the case of unitary electronic spin evolution (for a pure spin state), $\frac{dm_{\alpha}}{dt}(\{ v_{\alpha'}\}, \{\theta_{0,\alpha'}\})$ is given by Eq.~(\ref{eq:dmdt_sc}).
In this case, we can use the relation $\partial_\theta H_{\rm sc} = \partial_{\theta_0}H_{\rm sc}$ and  the semiclassical Schr\"{o}dinger equation (\ref{eq:shrodinger_sc}) to perform integration by parts with respect to $\theta_0$, and rewrite Eq.~(\ref{dmdt_phaseavg}) as
\begin{equation}
\label{dmdt_phaseavg2} \overline{\frac{dm_\alpha}{dt}}(\{ v_{\alpha}\}) = \frac{d}{dt}\left[\frac{-i}{\left(2\pi\right)^3}\oint d^3\theta_0\, \psi^{\dagger}\frac{\partial\psi}{\partial\theta_{0,\alpha}}\right].
\end{equation} 
The dependence on the Overhauser field magnitudes $\{ v_{\alpha}\}$ on the right hand side of Eq.~(\ref{dmdt_phaseavg2}) is implicitly incorporated in the evolution of the qubit state represented by $\psi$.

Intriguingly, Eq.~(\ref{dmdt_phaseavg2}) shows that the phase-averaged nuclear polarization rate can be expressed as a total time derivative.
Therefore, the time-integral over one cycle trivially gives
\begin{equation}
\overline{\delta m}_\alpha(\{ v_{\alpha}\}) = \left.\frac{-i}{\left(2\pi\right)^{3}}\oint d^3\theta_0\, \psi^{\dagger}\frac{\partial\psi}{\partial\theta_{0,\alpha}}\right|_{t_{i}}^{t_{i+1}},\label{eq:m_bar_v2}
\end{equation}
where 
\begin{equation*}
\left.\psi^\dagger \frac{\partial\psi}{\partial\theta_{0,\alpha}}\right|_{t_{i}}^{t_{i+1}} = \psi^\dagger \frac{\partial\psi}{\partial\theta_{0,\alpha}}(t_{i+1}) - \psi^\dagger \frac{\partial\psi}{\partial\theta_{0,\alpha}}(t_{i}).
\end{equation*}

 Remarkably, expression (\ref{eq:m_bar_v2}), which is a direct result of the unitary evolution of the electron spin states, depends only on the initial and final electronic states at times $t_i$ and $t_{i+1}$. 
In contrast, note that in the case of non-unitary electron spin evolution, as described for example by the master equation (\ref{eq:lindblad_eqution}), no such simple relation exists.
Therefore, to treat the noisy case we directly numerically integrate expression (\ref{eq:dmdt_rho_v1}) for the polarization rate, using solutions to the time-dependent master equation (\ref{eq:lindblad_eqution}).
Nonetheless, Eq.~(\ref{eq:m_bar_v2}) provides a very useful launching point for investigating the physical origins of the oscillatory DNP production rate.
We will proceed with the analysis for the case of unitary time evolution, and will later compare the results with those of the numerical simulations presented in Fig.~\ref{fig:DNP}.


\subsection{Rotating frame transformation}
We now make a unitary tranformation $R(t)$ (defined below) to a rotating frame in which the transformed state spinor $\tilde{\psi}(t) = R(t)\psi(t)$ evolves under a Hamiltonian with {\it real} entries.
The wave function $\tilde\psi(t)$ evolves according to 
\begin{equation}
\label{eq:eom_psi_d}i\frac{d}{dt}\tilde\psi = \tilde{H}(t)\tilde\psi,\ \ \tilde{H}(t) =  i\frac{dR}{dt}R^\dagger + RH_{\rm sc}(t)R^\dagger.
\end{equation}
All entries of $\tilde{H}(t)$ become real with the choice $R(t) = e^{i\phi(t)\sigma_z/2}$, with $\phi = \arg \left[v_{\rm SO} + v_{\rm HF}(t)\right]$, see Eq.~(\ref{eq:v_HF}):
\begin{equation} 
\label{eq:H_tilde}\tilde{H}(t) = 
\left(\begin{array}{cc}
\frac12\Delta(t) + \frac12\dot{\phi} & v_{\varphi}(\left\{ \theta(t)\right\} )\\
v_{\varphi}(\left\{ \theta(t)\right\} ) & -\frac12\Delta(t) - \frac12\dot{\phi}
\end{array}\right),
\end{equation}
with $v_{\phi}(t) = |v_{\rm HF}(t) + v_{\rm SO}|$.
The transformation $R(t)$ simply applies a relative phase to the singlet and triplet components of $\psi$, without changing the occupation probabilities. 







The dynamical part of the evolution, which occurs in the rotating frame, is described by the time evolution operator $U_d(t, t_0)$: $\tilde{\psi}(t) = U_d(t, t_0)\tilde{\psi(t_0)}$ with
\begin{equation}
U_d(t, t_0) = \mathcal{T}e^{-i\int_{t_0}^{t} \tilde{H}(t') dt'},
\end{equation}
where $\mathcal{T}$ is the time-ordering operator.
Given an initial {\it laboratory-frame} spinor $\psi(t_i)$ at the beginning of the $i$-th cycle, the initial state spinor for the following cycle can be written as 
\begin{equation}
\psi(t_{i+1}) = R(t_{i+1})U_d(t_{i+1},t_i)R^\dagger(t_i)\psi(t_i). 
\end{equation}
Inserting this expression into Eq.~(\ref{eq:m_bar_v2}), we obtain two contributions for the polarization.
The first, which we denote the ``geometrical contribution'' $\overline{\delta m}^{(g)}_{\alpha}$, arises from the $\theta$-derivatives of the rotating frame transformation $R(t)$ and its inverse: 
\begin{equation}
\overline{\delta m}^{(g)}_{\alpha} = \left.\frac{1}{2(2\pi)^{3}}\oint d^{3}\theta_{0}\left[\left(P_{T_{+}}-P_{S}\right)\frac{\partial\varphi}{\partial\theta_{0,\alpha}}\right]\right|_{t_i}^{t_{i+1}},\label{eq:delta_m_g}
\end{equation}
where $P_{T_{+}}=|\psi_{T_{+}}|^{2}$ and $P_{S}=|\psi_{S}|^{2}$
are the probabilities to find the electronic subsystem in the triplet and singlet states, respectively. 
In addition, the $\theta$-derivates of $U_d$ give rise to a ``dynamical contribution'' $\overline{\delta m}^{(d)}_{\alpha}$ to the polarization,
\begin{equation}
\overline{\delta m}^{(d)}_{\alpha} = \left.\frac{-i}{(2\pi)^{3}}\oint d^{3}\theta_{0}\,\tilde{\psi}^{\dagger}\frac{\partial\tilde{\psi}}{\partial\theta_{0,\alpha}}\right|_{t_{i}}^{t_{i+1}}.\label{eq:delta_m_d}
\end{equation}
The total change in polarization of species $\alpha$ in the $i$-th cycle is given by the sum of the two contributions, $\overline{\delta m}_{\alpha} = \overline{\delta m}^{(g)}_{\alpha}+\overline{\delta m}^{(d)}_{g}$. 




The time-dependent transformation $R(t)$ applied here is a generalization of the unitary rotation used to analyze the single-species problem in Ref.~\onlinecite{Rudner2010}.
There, the effects of nuclear Larmor precession during the sweeps and during the short waiting time $t_{\rm wait,1}$ near the $S$-$T_+$ avoided crossing were ignored.
Within this approximation, a symmetry argument was used to show that the dynamical contribution to the phase-averaged polarization change {\it vanished}.
Here we take into account multiple nuclear spin species as well as the effects of nuclear Larmor precession throughout the entire cycle.
Below we show that, in general, the dynamical contribution $\overline{\delta m}^{(d)}_\alpha$ does not vanish, and that it can lead to large contributions to the net polarization.

Note that the distinction between geometrical and dynamical contributions is arbitrary; under certain conditions where no polarization is expected, the two terms produce large but opposing contributions which cancel in the net polarization rate (see below).  
Nonetheless, we find this decomposition to be quite useful for analysis.


\subsection{The geometrical contribution to DNP\label{sub:The-geometrical-polarization}}

In order to understand the properties of the geometrical contribution to the polarization, we analyze Eq.~(\ref{eq:delta_m_g}) for the cases of one and two sweeps through the $S$-$T_{+}$ anti-crossing. 
We assume for simplicity that the system is initialized in the singlet state $\Ket{S}$ at the beginning of each cycle, which means $P_{S}(t=0)=1$. 

First we consider a cycle which involves a single sweep, which takes place over a time $t_{\rm sweep}$ which is much shorter than the nuclear Larmor precession period for any species. 
In this case we can neglect the Larmor precession and assume that the phases of the Overhauser fields are static: $\theta_{\alpha}(t)=\theta_{0,\alpha}$.
We note that, as shown for a single species in Ref.~\onlinecite{Rudner2010}, the dynamical contribution in Eq.~(\ref{eq:delta_m_d}) vanishes after averaging over all values of $\{\theta_{\alpha}\} $ due to the symmetry $\tilde{\psi}(\{\theta_{0,\alpha}\}) = \tilde{\psi}(\{-\theta_{0,\alpha}\})$.
The average polarization change results from the geometrical polarization alone, $\overline{\delta m}_{\alpha} = \overline{\delta m}^{(g)}_{\alpha}$, with
\begin{equation}
\overline{\delta m}^{(g)}_{\alpha} = \frac{1}{(2\pi)^{3}}\oint d^{3}\theta_0\, P_{T_{+}}(t_{\rm sweep},\{ \theta_{0,\alpha}\})\frac{\partial\varphi}{\partial\theta_{0,\alpha}}.\label{eq:m_g_P_t}
\end{equation}
According to Eq.~(\ref{eq:m_g_P_t}), the geometrical contribution is given by a weighted average over the electronic singlet-triplet spin flip probability. 
This transition probability depends on the phases $\{\theta_{0, \alpha}\}$, because the size of the coupling (i.e.~the size of the gap) at the $S$-$T_+$ avoided crossing is determined by the vector sum of the Overhauser fields (along with the effective spin-orbit field). 

Generally, the geometrical contribution to the polarization for a single sweep (with singlet initial state) 
is positive and has a magnitude corresponding to no more than one nuclear spin flip per sweep,
\begin{equation}
0\le\sum_{\alpha}\overline{\delta m}^{(g)}_{\alpha}\le1,\quad ({\rm single\ sweep}).\label{eq:m_g_g0_l1}
\end{equation}

Now consider the limit of an adiabatic sweep, where $\frac{d\Delta_c}{dt} \ll |v_{\varphi}(\{\theta_{\alpha}\})|^{2}$ for Overhauser field configuration $\{\theta_{\alpha}\}$, with fixed magnitudes $\{v_\alpha\}$. 
For some choices of Overhauser field magnitudes $\{v_\alpha\}$, there may be configurations of field orientations for which there is perfect destructive interference and the singlet-triplet coupling vanishes, $v_{\varphi}(\{\theta_{\alpha}\}) = 0$.
The following discussion about the adiabatic limit applies when the fixed values of $\{v_\alpha\}$ are chosen such that $v_{\varphi}(\{\theta_{\alpha}\})$ remains nonzero for all $\{\theta_{\alpha}\}$.

In the adiabatic limit, the electronic spins flip from the singlet state to the the triplet state with close to unit probability, $P_{T_{+}}(t_{\rm sweep})\simeq1$. 
Here Eq.~(\ref{eq:delta_m_g}) reduces to 
\begin{equation}
\label{eq:dmg_ad}\overline{\delta m}^{(g)}_{\alpha} \simeq \frac{1}{(2\pi)^{3}}\oint d^{3}\theta_0\,\frac{\partial\varphi}{\partial\theta_{0,\alpha}}.
\end{equation}
Generally, the geometric contribution in the adiabatic limit, Eq.~(\ref{eq:dmg_ad}), depends on the magnitudes of all three transverse Overhauser fields, $\{v_{\alpha}\}$, and on the strength of the spin-orbit coupling, $v_{\rm SO}$. 

In the case of only a single nuclear species, with Overhauser field $v_{\rm HF}=ve^{i\theta}$, we have $\varphi=\arg\left[ve^{i\theta}+v_{\rm SO}\right]$, and the polarization becomes\cite{Rudner2010}
\begin{equation*}
\overline{\delta m}^{(g)} = \frac{1}{(2\pi)}\oint d\theta\frac{\partial\varphi}{\partial\theta}.
\end{equation*}
The expression on the right hand side is a topological index associated with the winding of $v_\phi$ around the origin of the complex plane. 
The average polarization change in this case is therefore quantized: $\overline{\delta m}$ equals $1$ if $v>v_{\rm SO}$, and equals $0$ if $v<v_{\rm SO}$. 

In the case where all three species contribute substantially to the total transverse Overhauser field, the adiabatic limit for which Eq.~(\ref{eq:dmg_ad}) applies can only be reached for certain combinations of spin-orbit and Overhauser field amplitudes where the $S$-$T_+$ gap remains open for all $\{\theta_\alpha\}$. 
In the limit of spin-orbit dominated dynamics, $v_{\rm SO} > \sum_\alpha v_\alpha$, the average polarization transfer is quantized and equal to 0.
Also, in the hyperfine dominated limit, we find that the total change of nuclear polarization is quantized to 1, $\sum_{\alpha}\overline{\delta m}_{\alpha} = 1$, if and only if $\sum_\alpha{v_\alpha e^{i\theta_\alpha}}+v_{SO}\neq0$ for any choice of the three phases $\{\theta_\alpha\}$.
For other combinations of parameters, or for faster sweeps, the adiabatic limit is not reached and the average polarization transfer resulting from a single sweep is not quantized.

We now turn to the case of multiple sweeps. 
In the absence of nuclear Larmor precession, the results in Eqs.~(\ref{eq:m_g_P_t}) and (\ref{eq:m_g_g0_l1}) above still hold: the only important dynamical quantity is the probably $P_{T_{+}}$ to find the system in the triplet state at the end of the manipulation sequence.
The results are modified, however, when we take into account nuclear Larmor precession, which causes the Overhauser field phases $\{\theta_\alpha\}$ to change over time.

Consider first the simple case of a single nuclear species with Larmor angular frequency $\omega$, and a detuning protocol $\Delta(t)$ which includes two identical but opposite sweeps (one forward and one back) through the $S$-$T_{+}$ anti-crossing, separated by a waiting time $t_{\rm wait}$.
We assume that $t_{\rm wait}$ is comparable to the Larmor period $\frac{2\pi}{\omega}$, but that the two sweeps occur on timescales much shorter than the Larmor period. 
This problem was analyzed in Ref.~\onlinecite{Rudner2010}.
Here we review the result for comparison to the more general case.

During the first sweep, the Overhauser field phase $\theta$ remains approximately fixed at the value $\theta_{0}$.
In the second sweep, the phase $\theta$ is again constant, but fixed at a different value $\theta_{0}+\omega t_{\rm wait}$ due to precession during the waiting period.
As described in Sec.~\ref{sub:Non Hermitian part}, we assume that electron spin coherence is lost during the waiting time due to fast electrical noise.
We take $\Ket{S}$ to be the initial state of the first sweep.
Due to the fast singlet-triplet decoherence between sweeps, the initial condition for the return sweep is an incoherent mixture of $\Ket{S}$ and $\Ket{T_+}$, with probabilities determined by the spin flip probability $F(\theta_0)$ during the first sweep.
As shown in Ref.~\onlinecite{Rudner2010}, Eq.~(\ref{eq:delta_m_g}) for this case reduces to 
\begin{equation}
\overline{\delta m}^{(g)} = \frac{1}{2\pi}\oint d\theta_0\left[F_1\frac{\partial\varphi_1}{\partial\theta_{0}}+\left(1-2F_1\right)F_2\frac{\partial\varphi_2}{\partial\theta_{0}}\right],\label{eq:m_g_P_t-1}
\end{equation}
where $F_1 = F\left(\theta_{0}\right)$ is the $S$-$T_+$ transition probability in the first sweep, $\varphi_1=\varphi(\theta_{0})$ is the phase of the $S$-$T_+$ matrix element of $H_{\rm sc}$ during the first sweep, and $F_2 = F(\theta_{0}+\omega t_{\rm wait})$ and $\varphi_2=\varphi(\theta_{0}+\omega t_{\rm wait})$ are the corresponding quantities for the second sweep. 

As shown in Ref.~\onlinecite{Rudner2010}, Eq.~(\ref{eq:m_g_P_t-1}) leads to polarization with a sign that is reversed relative to that expected for a single sweep, in a large portion of the spin-orbit dominated regime.
This reversal means that, on average, more hyperfine-mediated transitions from $\Ket{T_{+}}$ to $\Ket{S}$ are expected than from $\Ket{S}$ to $\Ket{T_+}$. 
Such a situation can only arise due to the presence of spin-orbit coupling, which relaxes the conservation of spin angular momentum.

Thus we find that, for multiple sweeps, the geometrical contribution to DNP need not be positive; rather, it is limited to the interval $-1 \le \overline{\delta m}^{(g)}_\alpha \le 1$. 
For the special condition $\omega t_{\rm wait} = \pi$, the dynamical contribution to DNP $\overline{\delta m}^{(d)}$ vanishes as for the case of a single sweep: $\overline{\delta m} = \overline{\delta m}^{(g)}$. 
For other waiting times, however, the dynamical contribution is significant and must be taken into account.
We analyze the dynamical contribution in the following subsection. 

When more than one species is present, the situation is more complex.
Although Eq.~(\ref{eq:m_g_P_t-1}) generalizes naturally, the analysis is less straightforward.
More importantly, the condition $\omega t_{\rm wait} = n\pi$, which guaranteed cancellation of the dynamical contribution in the single species case, does not generalize to the multi-species case where each species has a different Larmor frequency.
Therefore both contributions must always be considered in order to correctly describe the net change in polarization.

\subsection{The dynamical contribution to DNP\label{sub:The-dynamical-polarization}}

The dynamical contribution to the phase-averaged DNP is defined in Eq.~(\ref{eq:delta_m_d}). 
As discussed above, the geometrical contribution depends on the electronic evolution only through the net changes in the singlet and triplet occupation probabilities. 
In contrast, the quantity $\tilde{\psi}^\dagger\frac{\partial\tilde{\psi}}{\partial\theta_{0,\alpha}}$ appearing in the dynamical contribution depends crucially on the \emph{phases} of $\tilde{\psi}$ at the initial and final times.

To gain further insight, we look at the dynamical contribution to the instantaneous phase-averaged polarization {\it rate}, Eq.~(\ref{dmdt_phaseavg}).
In the rotating frame, the dynamical contribution to $\frac{dm_\alpha}{dt}$ is given by the expression in Eq.~(\ref{eq:dmdt_sc}) with $H_{\rm sc}$ replaced by $\tilde{H}$. 
This gives
\begin{equation}
\overline{\frac{dm^{(d)}_{\alpha}}{dt}} = -\frac{1}{\left(2\pi\right)^{3}}\oint d^{3}{\theta}\ \tilde{\psi}^{\dagger}\frac{\partial \tilde{H}(t)}{\partial\theta_{\alpha}}\tilde{\psi}.\label{eq: dmd_dt}
\end{equation}

By construction, the Hamiltonian $\tilde{H}$ is a real, Hermitian $2\times 2$ matrix.
Thus for a given Overhauser field configuration $\{v_\alpha, \theta_{\alpha}(t)\}$, $\tilde{H}(t)$ has two instantaneous eigenvalues, $\tilde{E}_{\pm}(t)$, and two real, instantaneous eigenvectors, $\tilde\chi_{\pm}(t) = (\tilde\chi_{\pm,S},\tilde\chi_{\pm,T^{+}})^{T}$, with $\tilde\chi_{\pm,S(T)} = \tilde\chi_{\pm,S(T)}^{*}$. 
In terms of projectors onto these instantaneous eigenvectors, we write
\begin{equation}
\tilde{H}(t) = \tilde{E}_{+}\tilde\chi_{+}\tilde\chi_{+}^{\dagger} + \tilde{E}_{-}\tilde\chi_{-}\tilde\chi_{-}^{\dagger}.\label{eq:H_d_Phi}
\end{equation}
The probabilities to find the system in the upper and lower instantaneous eigenstates are given by $\tilde{P}_{\pm} = |\tilde\chi_{\pm}^{\dagger}\tilde\psi|^2$.
When $\dot{\phi}$ is small compared with $\Delta$ and/or $v_\phi$, the instantaneous eigenvalues $\tilde{E}_\pm$ are approximately equal to the eigenvalues $E_\pm$ of the laboratory frame Hamiltonian $H_{\rm sc}$, and $\tilde{P}_\pm \approx P_\pm$.
This condition is typically satisfied, since the nuclear Larmor frequencies at the field values used in experiments are small compared with the electronic hyperfine splitting for a random nuclear spin state.
Therefore, below we neglect the difference between these quantities.

We now use the form in Eq.~(\ref{eq:H_d_Phi}) to simplify the integrand in Eq.~(\ref{eq: dmd_dt}).
The derivative $\tilde{\psi}^{\dagger}\frac{\partial \tilde{H}}{\partial \theta_\alpha}\tilde{\psi}^{\dagger}$ has two types of contributions: the first type involves derivatives of the eigenvalues, $\frac{\partial E_\pm}{\partial \theta_\alpha}$, while the second type involves derivatives of the projectors, $\frac{\partial}{\partial \theta_\alpha}\left(\tilde\chi_\pm\tilde\chi^\dagger_\pm\right)$. 
The terms in which the derivative acts on the eigenvalues are 
diagonal in the instantaneous eigenbasis of $\tilde{H}$.
We thus call these terms the ``incoherent'' part of 
the polarization rate, since they are insensitive to coherence in the electronic spinor (in the $\tilde\chi_\pm$ basis).
On the other hand, the terms arising from the derivatives acting on the projectors are proportional to ${\rm Re}\left\{(\tilde{\psi}^\dagger\partial_{\theta_\alpha}\tilde{\chi}_\pm)\cdot(\tilde{\chi}_\pm^\dagger\tilde{\psi})\right\}$, and contribute {\it only} when the electronic spinor maintains coherence between $\tilde\chi_+$ and $\tilde\chi_-$.
Why is this so?
Noting that $\tilde\chi_+$ and $\tilde\chi_-$ are defined to be real and normalized, we have $\partial_{\theta_\alpha}\tilde\chi_+ \propto \tilde\chi_-$, and vice-versa. 
Hence the quantity ${\rm Re}\left\{(\tilde{\psi}^\dagger\partial_{\theta_\alpha}\tilde{\chi}_+)\cdot(\tilde{\chi}_+^\dagger\tilde{\psi})\right\}$ is proportional to ${\rm Re}\left\{(\tilde{\psi}^\dagger\tilde{\chi}_-)\cdot(\tilde{\chi}_+^\dagger\tilde{\psi}) \right\}$, which vanishes unless the electronic state spinor has nonzero components in both vectors $\tilde{\chi}_+$ and $\tilde\chi_-$.  
Furthermore, since the amplitudes $\tilde\chi_\pm^\dagger \tilde\psi$ carry phases which wind as $e^{-iE_\pm t}$, this ``coherent'' part of the polarization rate will oscillate at the frequency $(E_+ - E_-)$.


We are primarily interested in the situation where the waiting time $t_{\rm wait,1}$ is much longer than both the oscillation period $(E_+ - E_-)^{-1}$, and the timescale of electron spin dephasing, $\gamma^{-1}$, over which these oscillations are damped.
Here, the contribution of the oscillatory terms to $\overline{\delta m}_\alpha^{(d)}$ is expected to nearly average to zero.
In this case, the dynamical contribution to the phase-averaged polarization rate is dominated by the incoherent part: 
\begin{equation}
\label{eq:dmd_dt_1}\overline{\frac{dm^{(d)}_{\alpha}}{dt}} = -\frac{1}{\left(2\pi\right)^{3}}\oint d^{3}\theta\left(P_{+}\frac{\partial E_{+}}{\partial\theta_{\alpha}} + P_{-}\frac{\partial E_{-}}{\partial\theta_{\alpha}}\right).
\end{equation}
This approximation is supported by the numerical results shown in Fig.~\ref{fig:DNP}b, where the difference between the black and red curves is due to the contribution of the coherent part.
For shorter waiting times, when $\gamma t_{\rm wait,1} \lesssim 1$, $\overline{\delta m}_\alpha^{(d)}$ picks up an oscillatory dependence on $t_{\rm wait, 1}$, which is small compared to the overall background of the incoherent part (see Fig.~\ref{fig:osc_vs_t1}b).

Equation (\ref{eq:dmd_dt_1}) for the dynamical contribution to the polarization has a simple physical interpretation. 
The spin-orbit matrix element $v_{\rm SO}$ in $H_{\rm sc}$, see Eq.~(\ref{eq:H_sc}), comes with a particular, device-geometry-defined phase.
As a result, the presence of spin-orbit coupling breaks the symmetry of $\tilde{H}$ with respect to simultaneous rotations of all nuclear spins about the $z$-axis, $\theta_\alpha \rightarrow \theta_\alpha + \delta \theta$ for all $\alpha$.
Thus the energies of the instantaneous electron spin eigenstates depend on the {\it phases} of the transverse Overhauser fields, $E_{\pm} = E_{\pm}(\{\theta_\alpha\})$. 

The dynamical contribution to the polarization rate for species $\alpha$ for any particular configuration $\{\theta_{\alpha'}\}$ is given by the $\theta_{\alpha}$-derivatives of these phase-dependent energies. 
This relationship is natural, as $m_{\alpha}$ and $\theta_{\alpha}$ are canonical variables; 
the energy dispersion gives the polarization rate, just as the energy dispersion $\frac{dE}{dp}$ for a single particle gives its velocity. 
It follows that, for a particular Overhauser field realization $\{\theta_{\alpha'}\}$, a large polarization can be produced simply by holding the system near the $S$-$T_{+}$ anti-crossing for a significant period of time.
The typical size of the effect is greatest near the anti-crossing, where the dispersion of $E_\pm(\{\theta_{\alpha'}\})$ is greatest, while its sign is random and depends on the realization $\{\theta_{\alpha'}\}$.
Thus significant cancellations in the phase-averaged polarization rate may be expected, while large contributions could remain in the {\it variance} of the net polarization produced over one cycle\cite{Brataas2011}.

The picture given above is modified when nuclear Larmor precession is taken into account.
Larmor precession breaks the symmetry responsible for the cancellations, leading to large contributions in the phase-averaged polarization itself.
For demonstration, consider the situation where the detuning $\Delta$ is held at a fixed value 
close to the $S$-$T_{+}$ avoided crossing.
For simplicity, assume that only one nuclear species is present.
The dispersion relations $E_\pm(\theta)$ can be viewed as representing the energy bands of a particle in a periodic potential (see Fig.~\ref{fig5} below), with $\theta$ playing the role of the crystal momentum.
For our choice of gauge, in which the spin orbit matrix element $v_{\rm SO}$ is real, the band gap between $E_+$ and $E_-$ is largest for $\theta = 0$; here $|v_\phi(\theta = 0)| = v_{\rm SO} + v_{\rm HF}$.
For $\theta = \pi$, the gap is minimal: $|v_\phi(\theta = \pi)| = v_{\rm SO} - v_{\rm HF}$.

In the Bloch-particle analogy, nuclear Zeeman energy corresponds to a potential which varies linearly with the position $m$ (as is the case for an electron in a uniform electric field).
We thus draw an analogy between Larmor precession and the Bloch oscillations of electrons in an ultra-clean lattice.
The linear (time-independent) potential can be traded for a phase shift that grows linearly in time, i.e.~precession: $\theta(t) = \theta_0 + \omega t$.

Suppose that the transverse Overhauser field is initially oriented with phase 0, i.e.~$\theta_0 = 0$, and that the electronic system is initialized in the ground state represented by $\chi_-$ with energy $E_-$.
When the Larmor frequency is much smaller than the minimal gap, the electronic state tracks the ground state with instantaneous energy $E_-[\theta(t)]$. 
Over one Larmor period, the system uniformly samples the dispersion relation of the lower band.
During the first half-period of the Larmor precession, the electronic energy grows as the state slides up the dispersion curve from the band minimum. 
The electronic energy saturates as $\theta$ reaches $\pi$, where the band maximum occurs.
Just as with conventional Bloch oscillations, where a change in kinetic energy can be linked to a change in potential energy in the applied electric field, here the change to the energy of the electronic spin state must be equal and opposite to a change in the Zeeman energy of nuclear spins.
Therefore the nuclear polarization must change by an amount $\delta m$ which satisfies $\Delta E = \delta m\,\omega$, where $\Delta E = E_-(\theta = \pi) - E_-(\theta = 0)$.

The bandwidth $\Delta E$ depends on the detuning $\Delta$.
The maximum bandwidth is achieved at the anti-crossing, $\Delta = 0$, where $\Delta E = 2v_{\rm SO}$. 
In this case we find the maximal polarization which can be achieved over one half Larmor period,
\begin{equation}
\delta m_{\rm max}(\theta_0 = 0) = \frac{2v_{\rm SO}}{\omega}.\label{eq:deltam_vso}
\end{equation}
For times beyond half of a Larmor period, the electronic energy starts going down, and the polarization rate reverses sign.
Note that this result relies on the initial condition $\theta_0 = 0$.
For other initial conditions, different amounts of net polarization may be achieved.
In particular, for $\theta_0 = \pi$, the same argument applies but with the opposite sign: $\delta m_{\rm max}(\theta_0 = \pi) = -\delta m_{\rm max}(\theta_0 = 0)$.

Within the single species model described above, we now estimate the size of the effect for realistic experimental parameters. 
Taking typical values for a GaAs double dot system, we use $v_{\rm SO} = \frac{Ja}{\sqrt{2}\ell_{\rm SO}}\Amp{\Delta,S}{(0,2)S}$ (see appendix and Ref.~\onlinecite{Stepanenko2012}), with roughly estimated spin-orbit length $\ell_{\rm SO}=40\ \mu$m \cite{Zumbuhl2002, Meunier2007, Amasha2008}, dot separation $a = 0.1\ \mu$m, and interdot tunnel coupling $J \simeq 10\ \mu$eV. 
We assume that electron Zeeman energy is comparable to the tunnel coupling, such that $S$-$T_{+}$ anti-crossing occurs at a detuning where the (1,1) and (0,2) singlet states are strongly hybridized, $\Amp{\Delta,S}{(0,2)S} \approx \frac{1}{\sqrt{2}}$. 
For these values we find $v_{\rm SO} \simeq 2\times10^7$ $\rm rad/s$, (in units where $\hbar=1)$.
For the Larmor frequency $\omega$, we use the gyromagnetic ratio of $^{75}$As 
and a magnetic field $B=100$ mT, similar to the field value used in experiments\cite{Foletti2008}.
This gives $\omega = 4.6\times10^6$ rad/s. 
According to Eq.~(\ref{eq:deltam_vso}), we find $\delta m_{\max} \simeq 9$.
This result indicates that as many as {\it ten} nuclei may be flipped in a single sweep\cite{Brataas2011}.

We end this section with few comments.
First, the discussion of the dynamical contribution above seems to rely on quantum mechanical concepts such as energy band dispersion relations and Bloch oscillations.
However, the effect we described is actually 
just the semi-classical mutual precession dynamics discussed in Sec.~\ref{sub:perp Knight field}.
Due to the spin-orbit interaction, the electronic spins tend to preferentially maintain a Knight field component along the $x$-direction.
The magnitude of this $x$-component depends on the nuclear configuration during the sweep.  
In the average over all nuclear configurations, this Knight field leads to a net imbalance between nuclear spin precession into the positive and negative $z$-directions, thus leading to DNP. 

Second, note that in the case of three nuclear species, each with its own Larmor frequency, the dynamical contribution to the polarization is found by integrating Eq.~(\ref{eq:dmd_dt_1}) numerically in time, once for each species $\alpha$.
The simple energy conservation consideration which led to Eq.~(\ref{eq:deltam_vso}) in the case of only one species is not powerful enough to uniquely determine the net polarization transferred to all three species. 
Note that in the multi-species case angular momentum can be transferred from the electrons to the nuclei, and between nuclear species via the electrons.
The latter effect, which conserves the total $z$-projection of nuclear spin angular momentum, may change the net Overhauser field acting on the electrons spins due to the species-dependent and position-dependent hyperfine coupling constants. Thus, even in the absence of spin-orbit interactions, the transfer of angular momentum between different nuclei can lead a time-dependence of the net Overhauser field that is different from the behavior of the total spin polarization.

Finally, note that the maximal polarization transfer $\delta m_{\rm max}(\theta_0 = 0)$ in Eq.~(\ref{eq:deltam_vso}) is not achievable in to-date experimental setups, where only the phase-averaged DNP production rate can be detected. 
Even in this case, however, special DNP protocols can be chosen to pick out large pieces of the dynamical contribution which survive the averaging over all initial Overhauser field phases. 
As we discuss in the next subsection, the key is to use finite-rate sweeps to control the Overhauser-field-dependent occupation probabilities $P_{+}(\{\theta_\alpha\})$ and $P_{-}(\{\theta_\alpha\})$. 

\subsection{Origin of the oscillatory DNP response}
In this subsection we provide a physical picture to explain the oscillatory dependence of DNP on cycle time. 
As demonstrated in Fig.~\ref{fig:DNP}, the oscillatory effect is mainly due to the incoherent part of the dynamical contribution to the DNP, Eq.~(\ref{eq:dmd_dt_1}), which occurs during the waiting time at $\Delta_c=\Delta_{1}$. 
For further simplification, we now reduce Eq.~(\ref{eq:dmd_dt_1}) for the three-species phase-averaged polarization rate to an effective one-species model, for any selected species $\alpha$. 
First we define effective single-species populations $\bar{P}_{\pm,\alpha}(\theta_\alpha)$ and energies $\bar{E}_{\pm,\alpha}(\theta_\alpha)$ 
by averaging the populations $P_\pm(\{\theta_{\alpha'}\})$ and the energies $E_\pm(\{\theta_{\alpha'}\})$ over the Overhauser field orientations of the remaining two species, $\alpha' \neq \alpha$: 
\begin{eqnarray}
\bar{P}_{\pm,\alpha}(\theta_{\alpha}) & = & \frac{1}{\left(2\pi\right)^{2}}\oint_{\alpha'\neq\alpha}d^{2}\theta P_{\pm}\label{eq:Pbar}\\
\bar{E}_{\pm,\alpha}(\theta_{\alpha}) & = & \frac{1}{\left(2\pi\right)^{2}}\oint_{\alpha'\neq\alpha}d^{2}\theta E_{\pm}.\label{eq:Ebar}
\end{eqnarray}
We then compute
\begin{equation}
\frac{dm^{(d)}_{\alpha}}{dt}\approx-\frac{1}{(2\pi)}\oint d\theta_{\alpha} (2\bar{P}_{+,\alpha}-1)\frac{d\bar{E}_{+,\alpha}}{d\theta_{\alpha}},\label{eq:dm_dt_app_alpha1}
\end{equation}
where we have used $P_{-,\alpha} = 1 - P_{+,\alpha}$ and $E_{-,\alpha} = -E_{+,\alpha}$.
In splitting the integrals in this way, we ignore correlations between $P_\pm$ and $E_\pm$ in Eq.~(\ref{eq:dmd_dt_1}).
Numerical evidence indicates that neglecting these correlations has relatively little effect (see below).

In Fig.~\ref{fig5} we show $\bar{P}_{\pm\alpha}$ and $\bar{E}_{\pm\alpha}$, for the case of $\alpha=^{75}$As.
The parameters are the same as in the simulation used to produce Fig.~\ref{fig:DNP}.
In particular, the total cycle time is set to $t_{\rm cycle} = 8.33\ \mu$s, which corresponds to one of the local minima of $\delta B^z_{\rm Tot}(t_{\rm wait})$, (off scale in Fig.~\ref{fig:DNP}b).
 Panels a and b  depict the situation in the steady state, at the beginning and at the end of the waiting period near the anti-crossing (with $t_{\rm wait,1}=400\,\rm ns$). 
The vertical axes are used to show the energies $\bar{E}_{\pm,\alpha}$, while the shading depicts the corresponding populations (dark indicating high population, light indicating low population).

Near the avoided crossing, the energies $E_\pm(\{\theta_\alpha'\})$ are determined by the magnitude of the vector sum of the Overhauser fields of the three species, and of the effective spin-orbit field.
After averaging over the Overhauser field orientations for $^{69}$Ga and $^{71}$Ga, the averaged energy $\bar{E}_{\pm,{\rm As}}(\theta_{\rm As})$ displays a residual dependence on $\theta_{\rm As}$ as shown in Fig.~\ref{fig5}.

At the beginning of the waiting period, just after the sweep, the occupation probability $\bar{P}_{+,{\rm As}}$ of the upper branch is maximal for $\theta_{\rm As} \approx \pm\pi$ (see Fig.~\ref{fig5}a).
Why does this happen?
The occupation probability after the sweep depends on two factors: the transition probability during the sweep, and the steady state populations of the two states at the beginning of the cycle (just before the sweep).
Related to the first, note that the gap at the avoided crossing is smallest for $\theta_{\rm As} \approx \pi$, for any configuration of the remaining two Overhauser fields.
Here, 
 the spin-orbit field and the $^{75}$As Overhauser field point in opposite directions.
The small gap leads to an enhanced probability of non-adiabatic transitions during the detuning sweep.
Importantly, however, the steady state occupation probabilities, $\bar{P}_{\pm,{\rm As}}$, depend on the transition probabilities averaged over many cycles.
For a generic cycle time, which is not commensurate with the Larmor precession period (i.e.~$t_{\rm cycle} \neq 2\pi n/\omega_{\rm As}$), the orientation $\theta_{\rm As}$ is different at the beginning of each cycle.
As a result, the steady-state occupation probabilites are smeared out, showing a relatively weak dependence on $\theta_{\rm As}$.
However, for a nearly commensurate cycle time, $t_{\rm cycle}\approx 2\pi n/\omega_{\rm As}$, the Overhauser field orientation 
hardly changes from one cycle to the next.
In this case, variations in the transition probabilities 
compound over many cycles.
This allows a significant dependence of $\bar{P}_{\pm,{\rm As}}$ on the orientation $\theta_{\rm As}$ to persist in the steady state.

Through a simple argument based on energy exchange between the electronic and nuclear subsystems, we now demonstrate how the $\theta_{\rm As}$-dependent imbalance of occupation probabilities, $\bar{P}_{+,{\rm As}} - \bar{P}_{-,{\rm As}}$, 
leads to DNP.
During the waiting period, $\theta_{\rm As}$ changes linearly in time due to Larmor precession.  
Correspondingly, the level populations shift approximately adiabatically along the energy curves $E_{\pm, {\rm As}}$ (see Fig.~\ref{fig5}b).
For the steady state distribution $P_\pm(\{\theta_\alpha\}$) described above, 
the electronic spins gain energy on average through each waiting period 
(compare ``before'' and ``after'' cases, in Figs.~\ref{fig5} a and b).
Because the system is only weakly coupled to an environment throughout this process, most of this change in the electronic energy is transferred from the nuclear spin system.
Specifically, nuclear Zeeman energy compensates the change in energy of the electronic spins, implying a negative change in the nuclear polarization. 

If 
there were no relaxation, energy would be fully conserved within the combined electron-nuclear system.
The electronic energy gain would be maximal when $t_{\rm wait,1}$ is close to half of the nuclear Larmor period. 
However, the jump terms in Eq.~(\ref{eq:lindblad_eqution}) cause equilibration of the electronic spin state during the waiting period, via energy exchange with the environment. 
As a result, the waiting time for obtaining maximum contrast of the DNP oscillations (i.e.~maximum energy transfer to the nuclei) depends on the equilibration time, as shown in Fig.~\ref{fig:osc_vs_t1}.

\begin{figure}[t]
\includegraphics[width=3.0 in]{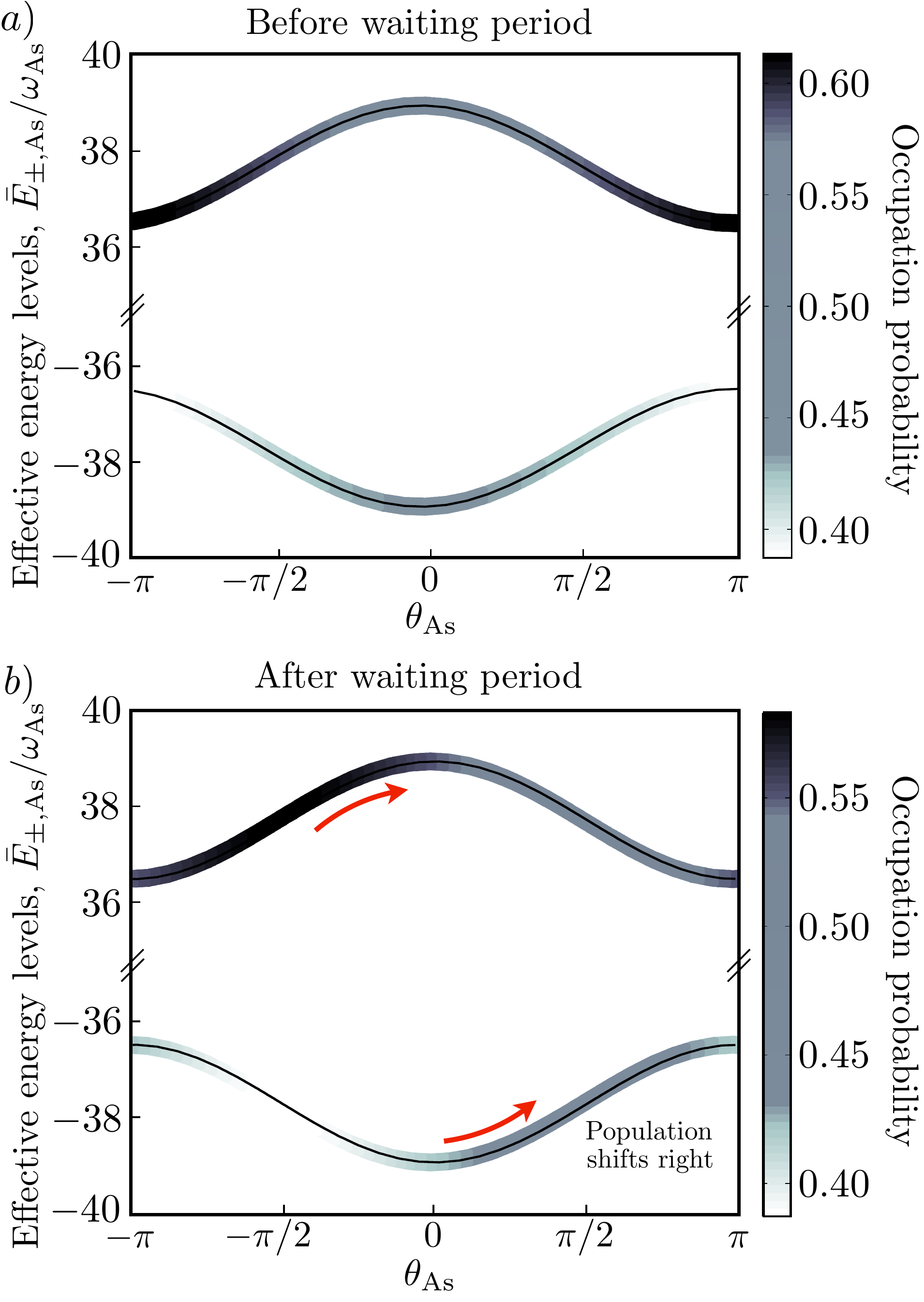}
\caption{Single-species model of DNP oscillations. The parameters are the same as in Fig.~\ref{fig:DNP}.
Focusing on the contribution of $^{75}$As, the effective energy levels $\bar{E}_{\pm,{\rm As}}$, Eq.~(\ref{eq:Ebar}), are plotted as a function of the orientation $\theta_{{\rm As}}$ of the $^{75}$As Overhauser field.
The corresponding averaged populations $\bar{P}_{\pm,{\rm As}}(\theta_{\rm As})$, Eq.~(\ref{eq:Pbar}), are represented by the line intensity (dark indicating high probability and light indicating low probability), shown before (panel a) and after (panel b) the waiting period.
Here the total cycle time is $t_{\rm{cycle}}=9.57\,\mu s$, which is very close to seven Larmor periods for $^{75}$As.
During the waiting period, $\theta_{\rm As}$ changes linearly in time due to Larmor precession, and the level populations adiabatically follow.  
For cycle times which are nearly commensurate with the nuclear precession period, differences in transition probabilities for the sweeps, which are controlled by the variations of the energy splitting with $\theta_{\rm As}$, compound over many cycles.
In the steady state, 
the electronic spins on average {\it gain} energy due to the precession during the waiting period. 
This energy is supplied by the nuclear Zeeman energy, implying DNP with a negative sign, as manifested in the sharp oscillatory dips. 
}
\label{fig5}
\end{figure}

The effective one-species model described above is only approximate, and so the accuracy of Eq.~(\ref{eq:dmd_dt_1}) must to be verified. 
In particular, recall that we explicity neglected correlations between the functions $P_\pm(\{\theta_\alpha\})$ and $E_\pm(\{\theta_\alpha\})$.  
Such correlations are expected, however, because $P_{+}(\{\theta_{\alpha}\})$ is controlled by the energy splitting $2E_{+}(\{\theta_{\alpha}\})$.
Consequently, it follows from Eq.~(\ref{eq:dmd_dt_1}) that the nuclear spins of different species interact and affect each other's polarization.
\begin{figure}[t]
\includegraphics[width=3.0 in]{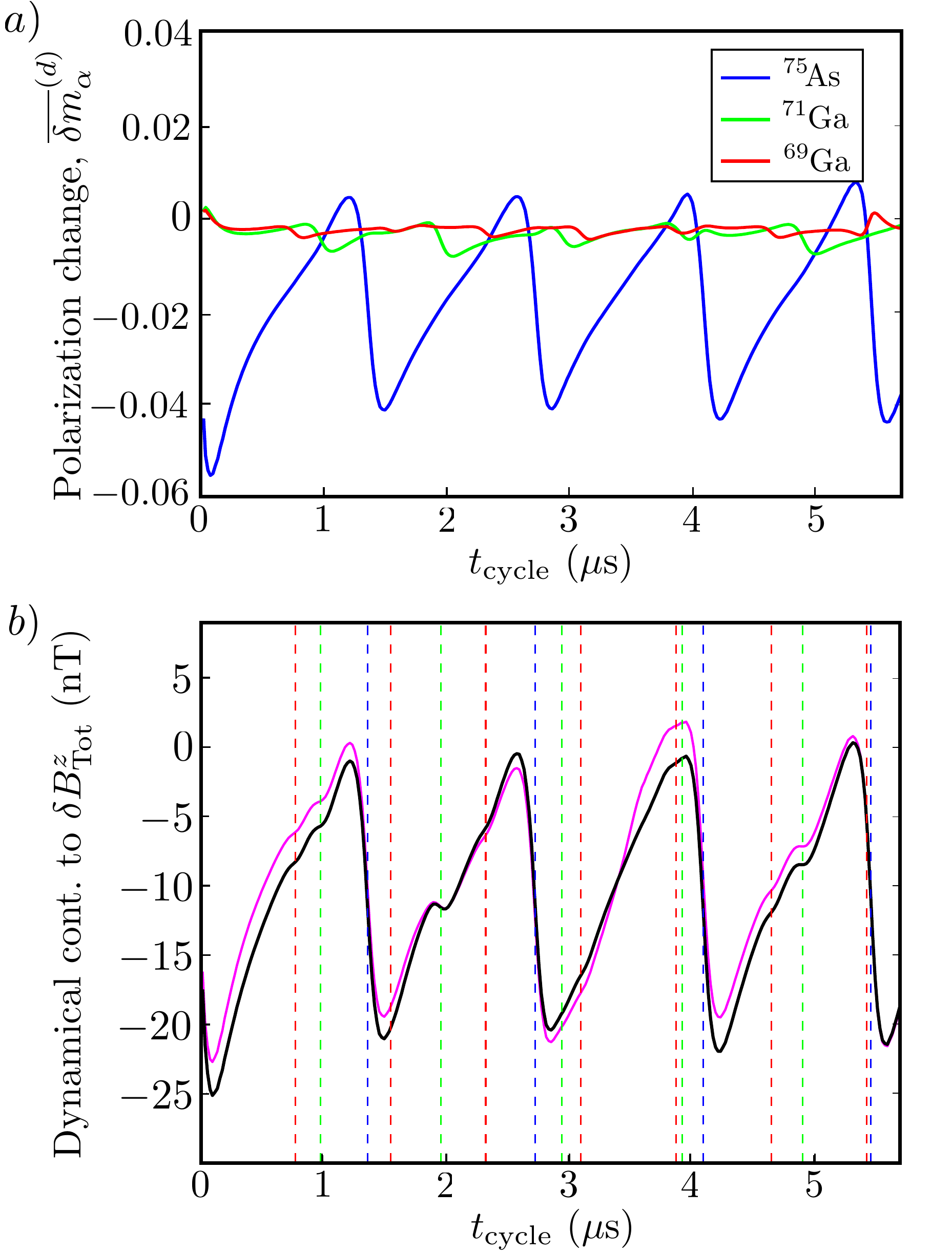}
\caption{
 Incoherent part of the phase-averaged steady state DNP per cycle, $\overline{\delta m}_\alpha^{(d)}$, calculated in the single-species approximation, Eq.~(\ref{eq:dm_dt_app_alpha1}). 
Protocol parameters are the same as in Fig.~\ref{fig:DNP}. a) Species-resolved contributions.
Compared with the results in Fig.~\ref{fig:DNP}a, here the behavior of each species is much more regular.
This difference arises because the single-species approximation neglects correlations which lead to inter-species polarization transfer. 
b) Change in the total Overhauser field per cycle $\delta B_{\rm Tot}^z\approx \frac1{N_d}\sum_{\alpha}\mathcal{A}_\alpha\overline{\delta m}_\alpha$. 
The black line shows the sum of the species-resolved results shown in the top panel. 
For comparison, the purple line shows the result obtained for the full three-species simulation, copied from Fig.~\ref{fig:DNP}b. 
\label{fig:DNP_1sp}}
\end{figure}

In Fig.~\ref{fig:DNP_1sp} we show a comparison between the numerically-obtained incoherent part of the dynamical contribution to DNP for the full three-species model, Eq.~(\ref{eq:dmd_dt_1}), and the corresponding result from the approximate single-species model, Eq.~(\ref{eq:dm_dt_app_alpha1}).
We see that the two curves are nearly identical. 
This might appear to be an indication that the oscillatory behavior primarily results from the independent coupling of each species to the electron spins, with little influence coming from the other species. 
However, when we compare the DNP production rate for each nuclear species, Figs.~\ref{fig:DNP}a and \ref{fig:DNP_1sp}a, we note that the single species approximation misses the polarization transfer from $^{75}$As to $^{69}$Ga, which is observed in the full three-species simulation for $t_{\rm cycle} \approx \frac{2\pi}{\omega_{2}-\omega_{1}}$.
This difference results from the neglected correlations between $E_{+}$ and $P_{+}$ in the $\left(\theta_{1},\theta_{2}\right)$ subspace (here $\theta_{1}$ is the direction of the $^{75}$As Overhauser field, and $\theta_{2}$ is the direction of the $^{71}$Ga Overhauser field).
Interestingly, despite the fact that the species-resolved DNP contributions are quite different for the two calculations, when we plot the {\it net} Overhauser field in Fig.~\ref{fig:DNP_1sp}b 
the two curves are almost identical. 
This result is actually a coincidence, arising from the fact that the microscopic hyperfine coupling constants $\mathcal{A}_\alpha$ for $^{75}$As and $^{69}$Ga are quite similar (1.92 T and 1.84 T, respectively).
As a result, polarization transfer between these two species hardly affects the total Overhauser field induced on the electron.

\section{Summary and Conclusions}\label{Discussion}

In this paper we analyzed the dynamical nuclear polarization (DNP) produced by repeated sweeps through the $S$-$T_+$ avoided crossing of a two-electron double quantum dot in an applied Zeeman field.
Coupling between the singlet and triplet states arises from a combination of the spin-orbit interaction and the hyperfine interaction with nuclear spins of the host crystal.
DNP is produced when angular momentum is transfered from the electron spins to the nuclear spins and vice-versa.

Interestingly, because the spin-orbit interaction breaks the separate conservation laws for spin and orbital angular momentum, large changes in the net nuclear polarization can result from the interaction between the nuclear spin subsystem and a single pair of electron spins, even if electrons are not reloaded between sweeps.
The sign of DNP is also unrestricted, and depends on the relative orientations of the electronic and nuclear transverse spin components. 
Importantly, under appropriate conditions we find that large contributions to DNP survive averaging over all initial orientations of the random nuclear spin state.

We adopted  a semi-classical treatment of the nuclear spins, in which the operator-valued transverse Overhauser field components are replaced by complex numbers with real and imaginary parts corresponding to the $x$ and $y$ field components.
Electron spin dynamics within the $S$-$T_+$ subspace are governed by a $2\times 2$ semiclassical Hamiltonian, with a complex off-diagonal matrix element $v_{\rm SO} + \sum_\alpha v_\alpha e^{i\theta_\alpha}$ that describes the combined action of the spin-orbit and hyperfine couplings.
This off-diagonal term is time-dependent, owing to nuclear Larmor precession in the external field: $\theta_\alpha(t) = \theta_{0,\alpha} + \omega_\alpha t$. 

DNP results from nuclear spin precession around the transverse components of the Knight field produced by the electron spins via the hyperfine interaction.
When the double-dot detuning is set close to the $S$-$T_+$ degeneracy point, the electron spins maintain transverse Knight field components due to the spin-orbit and Overhauser terms which dominate the semiclassical Hamiltonian. 
Thus the semi-classical equations of motion for the nuclear spins are nonlinear: the nuclear state affects the evolution of the electronic state, which in turn controls nuclear precession.

We numerically integrated the semi-classical equations of motion for a protocol similar to the one used in the experiments in Ref.~\onlinecite{Foletti2008}.
The protocol is periodic, with fast sweeps between a large positive detuning $\Delta_0$, and a small detuning $\Delta_1$ near the $S$-$T_+$ anti-crossing.
The sweeps are separated by two long waiting periods, $t_{\rm wait, 1}$ at detuning $\Delta_1$ and $t_{\rm wait, 2}$ at detuning $\Delta_0$. 
Importantly, there is no re-initialization of the electronic state between cycles.
However, in the model we allowed dephasing of the electron spin state during the cycle, e.g.~due to 
electronic noise which couples to the charge degrees of freedom, and also included a probability for the triplet state to relax to the singlet state during the waiting period $t_{\rm wait,2}$. We did not include any direct effects of the relaxation on the nuclear system, as would be produced if spin relaxation is mediated by hyperfine processes.  Here we assume that the relaxation at large positive detuning is dominated either by electronic exchange to a nearby reservoir (i.e.~a lead), or by spin-orbit-mediated coupling to phonons. 

The numerical results, Fig.~\ref{fig:DNP}, show an oscillatory dependence of the steady state DNP production rate on total cycle time, which is controlled by varying $t_{\rm wait,2}$ with all other parameters held constant.
Similar to the experimental observation, the phase-averaged DNP per cycle displays dips when the total cycle time, $t_{\rm cycle} = t_{\rm sweep} + t_{\rm wait, 1} + t_{\rm wait, 2}$, is commensurate with the Larmor period of any one of the three species. 
Furthermore, the oscillation amplitude shows strong dependence on the waiting time near the anti-crossing, $t_{\rm wait,1}$, and is maximal for 
waiting times which are long relative to the electron spin coherence timescale (Fig.~\ref{fig:osc_vs_t1}).

In order to gain intuition from these results, we analyzed the semi-classical equations of motion, by moving to a convenient rotating frame where the phase of the singlet-triplet matrix element is constant in time.
We focused on the ``phase-averaged'' change in DNP due to one cycle of the DNP protocol, meaning that we averaged the DNP produced by one cycle over all initial transverse Overhauser field orientations, with fixed magnitudes $\{v_\alpha\}$.
We then examined a ``dynamical contribution,'' Eq.~(\ref{eq:delta_m_d}), which is produced by the dynamics under the Hamiltonian in the rotating frame, and a ``geometrical contribution,'' Eq.~(\ref{eq:delta_m_g}), which comes from the rotating-frame transformation itself.

The geometrical contribution $\overline{\delta m}_\alpha^{(g)}$ is a function of the probabilities for the electrons to occupy the singlet and triplet states before and after a given cycle. 
In the absence of spin-orbit coupling, the geometrical contribution is exactly one nuclear spin flip per electron spin flip, due to conservation of spin angular momentum. 
In the presence of spin-orbit coupling, the geometrical contribution is quantized (with a value that can be 0 or 1) in the case of an adiabatic sweep through the $S$-$T_+$ anti-crossing, where the electron spin flip probability is unity for any configuration of Overhauser field orientations $\{\theta_\alpha\}$.
In a more general, non-adiabatic sweep, the geometrical contribution can take any value between -1 and 1. The geometrical contribution dominates the phase-averaged DNP when the Larmor precession of the nuclear spins is much slower than the overall cycle time.

The main focus of our analysis is 
the dynamical contribution to the averaged DNP 
in the case of finite Larmor precession periods. 
Physically, this contribution originates from the fact that the electron spins maintain an average component in the direction of the spin-orbit field for detunings near the  $S$-$T_+$ anti-crossing.
The corresponding component of the Knight field acting on the nuclear spins is transverse to the external field.
Nuclear precession around the total (external plus Knight field) therefore leads to changes of the $z$-component of total nuclear polarization.
Importantly, 
sweeping through the anti-crossing creates a correlation between the electron and nuclear spin orientations, 
which allows this polarization to survive the averaging over all initial nuclear spins orientations.  
Moreover, the simulation clearly shows that the dynamical contribution to DNP continues to grow while the system 
is held near the $S$-$T_{+}$ anti-crossing, 
long after a complete loss of the coherence of the electron spins.

To better understand this new effect, we identified an ``incoherent part'' of the dynamical contribution to DNP, Eq.~(\ref{eq:dmd_dt_1}), that dominates during the waiting period near the crossing, and that is 
insensitive to the electron spin coherence. 
We understand this term through an analogy between the  function $E_\pm(\{ \theta_{\alpha'}(t)\})$, i.e.~the Overhauser field dependent electron spin energy levels, and the energy bands of an electron in a periodic potential. 
In this analogy, the variables $\{ \theta_{\alpha'}(t)\}$ play the roles of the conjugate momenta to the polarization variables $\{m_\alpha\}$. 
The dispersion relations $\frac{\partial E_\pm}{\partial \theta_\alpha}$ give the corresponding velocities, or the polarization rates $\{\dot{m}_\alpha\}$. 
The Larmor precession due to the Zeemann coupling is mapped onto the Bloch oscillations exhibited by the Bloch-electron in a uniform electric field (linear potential). 
The DNP arising during the waiting period near the anti-crossing can then be understood by a simple energy consideration argument.
Slow Larmor precession causes an adiabatic change in the energy levels of the electronic spin system; this change in energy must be extracted from the Zeeman energy of the nuclear subsystem. 
Hence changes in the electronic energy are linked to changes in nuclear polarization.
We showed numerically that, even though the problem involves three nuclear species, an effective model based on a single nuclear species and spin-orbit field produces qualitatively similar oscillatory behavior of the DNP. 
   
With these insights, the oscillatory behavior of the steady-state DNP as a function of the total cycle time in  Fig.~\ref{fig:DNP} can be explained as follows.  
One important consequence of Eq.~(\ref{eq:dmd_dt_1}) is that, in order to obtain a non-zero phase-averaged DNP, the system must maintain steady state populations $P_{\pm}(\{\theta_{\alpha}\})$ of the upper and lower instantaneous eigenstates which depend in a non-trivial way on the Overhauser field orientations. 
This situation is indeed achieved by repeatedly sweeping through the $S$-$T_+$ anti-crossing with a sweep-rate comparable to the typical value of the gap squared, $|v_{\rm SO} + \sum_\alpha v_\alpha e^{i\theta_\alpha}|^2$. 
Because the gap is sensitive to the Overhauser field configuration $\{\theta_{\alpha}(t)\}$, the transition probabilities between the two electron spin energy levels, and hence the steady state populations, in general depend on $\{\theta_{\alpha}(t)\}$.
However, in order for the steady-state probability to reflect this dependence, the total cycle time must be close to commensurate with the Larmor period of a given nuclear species $\alpha$ such that the value  of $\theta_\alpha$  repeats itself from cycle to cycle, and with it the transition probabilities. 
The oscillatory response is strongest when a) the sweeps back and forth through the $S$-$T_+$ avoided crossing are fast, such that the system is far from the adiabatic regime, b) the probability $\Lambda$ for the electronic state to decay/reload to a singlet at the end of each cycle is small, and c) the waiting time near the $S$-$T_{+}$ anti-crossing, $t_{\rm wait, 1}$, is comparable to half the Larmor period.
   
In the experiment in Ref.~\onlinecite{Foletti2008}, many paramters such as the detuning at the waiting point, $\Delta_{1}$, the dephasing rate $\gamma$, and the decay probability $\Lambda$ are only known through rough estimatees. 
Therefore we can only make a qualitative comparison between the experimental results and the predictions of the theory.
Within the accuracy to which the experimental parameters are known, it appears reasonable that the three criteria listed above for producing DNP oscillations were met in the experiment.
The optimal waiting time for producing the maximum oscillation amplitude was found experimentally to be $t_{\rm wait, 1} = 100\ \mu$ s. As we show in Fig.~\ref{fig:osc_vs_t1}, this result can be explained if we assume a short dephasing time of the electron spin coherence, $\gamma^{-1}\sim10$ ns.

The theory presented here can be quantitatively confronted by future experiments with various DNP protocols. For example, it would be instructive to measure the DNP for the reload case, $\Lambda=1$, and its dependence on parameters such as $t_{\rm wait, 1}$ and $\gamma$.
Another interesting avenue to explore is the dependence on external magnetic field direction within the plane of the 2DEG.
The field orientation controls the value of the spin-orbit matrix element $v_{\rm SO}$, and thus gives a useful knob for investigating the interplay between spin-orbit and hyperfine couplings in confined geometries.
\begin{acknowledgments}
We thank Leonid Levitov for helpful discussions throughout this work.  This research was supported in part by NSF grant DMR-0908070 and by the Office of the Director of National Intelligence, Intelligence Advanced Research Projects  Activity (IARPA), through the Army Research Office grant W911NF-12-1-0354. I.N. was partly supported by the Tel Aviv University Center for Nanoscience and Nanotechnology.
\end{acknowledgments}
\appendix
\section*{Appendix A - Model f\label{sec:Appendix-A--}or the spin-orbit effective
interaction in double quantum dot}

In this appendix we 
calculate the matrix elements of the spin-orbit coupling Hamiltonian within the  $S$-$T_+$ subspace of a two-electron double quantum dot formed in a GaAs 2D electron gas (see Eq.~(\ref{eq:vso}) of the main text). 
Ignoring the coupling to nuclei, the double quantum dot Hamiltonian projected onto the $S$-$T_{+}$ subspace consists of three terms:
\begin{equation}
\label{eq:Happ}
H=H_{\rm orb}+H_{B}+H_{\rm SO}.\end{equation}

The first is the orbital part, which we take in our model to be

\begin{equation}
\label{eq:Horb}
H_{\rm orb}=\sum_{i=1,2}{\left[\frac{\Pi_{i}^{2}}{2m_{e}}+V_{\Delta}(\vec{r}_{i})\right]}+U(\vec{r}_{1},\vec{r}_{2}),
\end{equation}
where the index $i$ labels the two electrons, $\vec{r}_{i}=\left(x_i,y_i\right)$ 
is the position operator for electron $i$ (within the plane of the 2DEG),
 and $\vec{\Pi}_{i}=\vec{p}_i-e\vec{A}\left(\vec{r}_i\right)$  is the corresponding two dimensional covariant momentum vectors.  Here $\vec{p}_{i}=\left(p_{x,i},p_{y,i}\right)$ is the canonical momentum and $\vec{A}\left(\vec{r}\right)$ the vector potential. 
In Eq.~(\ref{eq:Horb}), $V_{\Delta}(\vec{r})$ represents the confining potential
that creates the double dot system, i.e.~it describes two separated potential wells, 
with a 
potential bias between the wells that depends on the detuning $\Delta$.  
Finally, $U(\vec{r}_1,\vec{r}_2)$ is the Coulomb interaction between the electrons. 
We assume that the interaction potential is symmetric to exchange transformation,
$U(\vec{r}_{1},\vec{r}_{2})=U(\vec{r}_{2},\vec{r}_{1})$, but otherwise leave its form completely general. 

The second term in Eq.~(\ref{eq:Happ}) is the Zeeman coupling
\begin{equation}
\label{eq:HBapp}
H_{B}=g^{*}\mu_{B}\vec{B}\cdot\left(\vec{S}_{1}+\vec{S}_{2}\right),
\end{equation}
where $\vec{S}_{1,2}$ are the spin vector operators of electrons 1 and 2 and $\vec{B}=B\vec{z}$ is the external magnetic field. We keep the  direction $\vec{z}$ of  the magnetic field  general through the derivation. At the end we discuss two cases, one in which $\vec{z}$
points perpendicular to the 2DEG,  and the other in which it points in some direction in the 2DEG plane.

The third term in Eq.~(\ref{eq:Happ}) represents the spin orbit coupling.
We include both Rashba-type and linear Dresselhaus contributions, which are expected to be important 
at low electron densities in the 2DEG \cite{Dresselhaus1955,Bychkov1983},

\begin{equation}
H_{\rm SO}=2\sum_{i=1,2}\alpha\left(\Pi_{i}^{y'}S_{i}^{x'}-\Pi_{i}^{x'}S_{i}^{y'}\right)+\beta\left(\Pi_{i}^{y'}S_{i}^{y'}-\Pi_{i}^{x'}S_{i}^{x'}\right),\label{eq: H_SO}\end{equation}
with $\alpha$ and $\beta$ the Rashba and Dresselhaus coefficients, respectively.
The $x',y'$ axes are oriented in the crystallographic directions [1 0 0] and [0 1 0] in the 2DEG plane (here we assume that the growth direction is [0 0 1]).

We first discuss the energy levels of the Hamiltonian $H_0=H_{\rm orb}+H_B$, i.e.~with no spin-orbit interaction. For the orbital part we assume a tight binding model, in which each of the two wells hosts a single bound state, while all other states are much higher in energy. We further assume a situation where the coupling between the wells is weak, such that the change of the wave function of the bound state in one well due to the presence of the other well can be neglected. However the bound states of the two wells do in fact overlap, which gives rise to 
electron tunneling between the two wells. 

We focus on the regime where the potential $V_\Delta$ is tuned such that the lowest energy state lies in the subspace formed by three states. The first is the singlet state $\left|(0,2)S\right\rangle $, in which the two electrons are in the right dot:  
\[
\left\langle \vec{r}_{1},\vec{r}_{2}\right.\left|(0,2)S\right\rangle =\psi_{(0,2)}^{S}(\vec{r}_{1},\vec{r}_{2})\left|S\right\rangle ,\]
where $\left|S\right\rangle$ is the spin singlet state and the orbital wave function  $\psi_{(0,2)}^{S}(\vec{r}_{1},\vec{r}_{2})$  has non-zero amplitude only when $\vec{r}_{1}$ and $\vec{r}_{2}$ are both positioned in the right dot. 
The two other states both have a (1,1) orbital wave function, where there is one electron in each dot. 
First, there is the singlet $\left|(1,1)S\right\rangle $:
\begin{equation}
\left\langle \vec{r}_{1},\vec{r}_{2}\right.\left|(1,1)S\right\rangle =\psi_{(1,1)}^{S}(\vec{r}_{1},\vec{r}_{2})\otimes\left|S\right\rangle ,\label{eq:11S}\end{equation}
whose orbital wave function $\psi_{(1,1)}^{S}(\vec{r}_{1},\vec{r}_{2})$ is symmetric under exchange of $\vec{r}_1$ and $\vec{r}_2$.  
Due to the spatial separation between the two dots, we assume that it  can be written as 
\[
\psi_{(1,1)}^{S}(\vec{r}_{1},\vec{r}_{2})=\frac{1}{\sqrt{2}}\left[\chi_{L}\left(\vec{r}_{1}\right)\chi_{R}\left(\vec{r}_{2}\right)+\chi_{L}\left(\vec{r}_{2}\right)\chi_{R}\left(\vec{r}_{1}\right)\right],\]
where $\chi_{L(R)}\left(\vec{r}\right)$ is a single particle wave function that is
localized on the left (right) dot. 
The other state is the triplet state $\left|(1,1)T_+\right\rangle$.
We assume that the corresponding orbital wave function is composed from the same single-particle orbitals as $|(1,1)S\rangle$, now combined in an anti-symmetric combination:
\begin{equation}
\left\langle \vec{r}_{1},\vec{r}_{2}\right.\left|(1,1)T_{+}\right\rangle =\psi_{(1,1)}^{A}(\vec{r}_{1},\vec{r}_{2})\otimes\left|T_{+}\right\rangle ,\label{eq:11T}\end{equation}
with
\[
\psi_{(1,1)}^{A}(\vec{r}_{1},\vec{r}_{2})=\frac{1}{\sqrt{2}}\left[\chi_{R}\left(\vec{r}_{1}\right)\chi_{L}\left(\vec{r}_{2}\right)-\chi_{L}\left(\vec{r}_{1}\right)\chi_{R}\left(\vec{r}_{2}\right)\right].\]
 
 The above tight binding model leads to an anti-crossing of the states $(0,2)S$ and $(1,1)S$, as one tunes the parameter $\Delta$,  which controls the inter-well bias potential. 
The anti-crossing is due to the electron tunneling between the wells, which is independent of spin, and therefore mixes only the states $\left|(1,1)S\right\rangle $ and $\left|(0,2)S\right\rangle$. 
When the Hamiltonian $H_0$ is projected onto the subspace spanned by the $(1,1)T_+$ state and these two singlet states, we assume that it takes the 
form 
\begin{equation}
H_S = \left(\begin{array}{ccc}
E_0-|g^*\mu_BB| & 0 & 0 \\
0 & E_0 & J \\
 0 & J & E_0-\epsilon \end{array}\right),\label{eq:H_S}
\end{equation}
where $E_0$ is the orbital energy of the $(1,1)$ state, $\epsilon$ the diabatic energy difference between the $(0,2)S$ and $(1,1)S$ singlet states, and the off-diagonal element $J$ describes 
electron tunneling.  We assume that $\epsilon\left(\Delta,B\right)$ is a monotonically increasing function of the detuning parameter $\Delta$, at each value of $B$, and we also allow that $J$ may also depend on gate voltages and magnetic field strength, $J = J(\Delta,B)$.

Note that we take the orbital contribution to the energy of the state $\left|(1,1)T_{+}\right\rangle$ to be the same as that of the singlet state $\left|(1,1)S\right\rangle$, in the absence of tunneling: 
 $E_{T_+} +|g^*\mu_BB| = E_{0}$.  
As we now explain, this approximation is consistent with the earlier approximation that we consider only a single orbital level in each dot. 
The difference of the orbital energy of the triplet state from $E_0$ arises from the exchange energy associated with the combination of the Coulomb interation between electrons and the overlap of the wave functions $\chi_L$ and $\chi_R$ under the barrier.  
Importantly, the tunneling amplitude $J$ is also proportional to this overlap, and to the ``attempt frequency'' which is given by the single-well orbital level spacing itself.
Thus when we work in the limit where both tunneling and electron-electron interactions are small compared with the single well level spacing, 
the overlap is small and we restrict 
the exchange term to be parametrically smaller than the tunneling term.
Here we keep only the zero-order contribution to the orbital energy in the small ratio of exchange energy over tunneling.
 
The corresponding lowest energy spin-singlet state of $H_S$ (and of $H_0$) is a superposition of $\left|(1,1)S\right\rangle $ and $\left|(0,2)S\right\rangle$, 
\begin{equation}
\label{eq:LowestS}
\left|\Delta,S\right\rangle =a\left(\Delta \right)\left|(1,1)S\right\rangle +b\left(\Delta\right)\left|(0,2)S\right\rangle,
\end{equation}
where the coefficients $a(\Delta)$ and $b(\Delta)$ describe the hybridization of the states. 
We denote the corresponding lowest eigenvalue $E_S$.
Note that, for any non-zero Zeeman splitting $g^{*}\mu_{B}B$, the energy $E_S$ of the lower energy singlet 
crosses that of $\left|(1,1)T_{+}\right\rangle$ for some value of the gate voltages which control the detuning; we define this point to be $\Delta=0$.

We now consider the perturbation $H_{\rm SO}.$  
For weak enough spin-obit coupling, we can neglect the contribution of the higher-energy singlet state, and consider only the two-fold low energy subspace spanned by the states $\left|\Delta,S\right\rangle$ and $\left|(1,1)T_{+}\right\rangle$, with unperturbed energies  $E_{S}$ and $E_{T_{+}}$, respectively.
In this approximation, we assume that the spin-orbit term $H_{\rm SO}$ has matrix elements which are much smaller than both the singlet energy splitting $J$  and the Zeeman splitting. 

We project $H_{\rm SO}$ onto the $\left\{\left|\Delta,S\right\rangle, \left|(1,1)T_{+}\right\rangle\right\}$ subspace in the following way. We first note that all the terms in $H_{\rm SO}$ are linear in the components of the momentum, $\Pi_i^m$, $m \in\left\{x,y\right\}$. 
These momentum components can be conveniently expressed in terms of commutators involving the {\it unperturbed} orbital Hamiltonian 
\begin{equation}
\Pi^m_i=-im_{e}[r^m_i,H_{\rm orb}]
\label{eq: p}.\end{equation}
Substituting this result into the spin-orbit Hamiltonian, Eq.~(\ref{eq: H_SO}), we obtain the spin-orbit coupling matrix elements in terms of matrix elements of the commutators $[r^m_i,H_{\rm orb}]$.

For evaluating the spin-orbit matrix elements, We use  Eqs.~(\ref{eq: H_SO}) and (\ref{eq: p}) together with the fact that the states $\left|\Delta,S\right\rangle$ and $\left|(1,1)T_{+}\right\rangle$ are eigenstates of the orbital Hamiltonian $H_{\rm orb}$, with eigenvalues $E_S$ and $E_0$, respectively. We find that the diagonal matrix elements of 
$\Pi_{j}^{m}$ in the $\left|\Delta,S\right\rangle$-$\left|(1,1)T_{+}\right\rangle$ basis vanish, which leads to
\begin{widetext}
\begin{equation}
\left\langle (1,1)T_{+}\right|H_{\rm SO}\left|(1,1)T_+\right\rangle=\left\langle \Delta,S\right|H_{\rm SO}\left|\Delta,S\right\rangle=0.
\end{equation}
This result is expected, as the states carry no net momentum. 
For the off-diagonal matrix element we find
\begin{eqnarray}
v_{\rm SO}&=&\left\langle (1,1)T_{+}\right|H_{SO}\left|\Delta,S\right\rangle   \nonumber\\
&= & -i\sum_{i=1,2}\sum_{m,n}2\kappa_{mn}\left\langle (1,1)T_{+}\right|S_{i}^{n}\left(r_{i}^{m}H_{orb}-H_{orb}r_{i}^{m}\right)\left|\Delta,S\right\rangle \nonumber \\
 & = & -i\left(E_{S}-E_{0}\right)\sum_{i=1,2}\sum_{m,n}2\kappa_{mn}\left\langle (1,1)T_{+}\right|S_{i}^{n}r_{i}^{m}\left|\Delta,S\right\rangle, \label{eq:11T_HSO_S}\end{eqnarray}
where the spin-orbit tensor $\kappa_{mn}$ is defined as $\kappa_{xx}=-\kappa_{yy}=-m_e\beta$ and $\kappa_{xy}=-\kappa_{yx}=-m_e\alpha$. 

We now insert Eqs.~(\ref{eq:11S}) and (\ref{eq:11T}) into Eq.~(\ref{eq:11T_HSO_S}), and use the fact that $r_{i}^{n}$ is a {\it local} operator, which does not couple states in which the electron $i$ is in different wells. This gives
\begin{equation}
v_{\rm SO} =\frac{i}{\sqrt{2}}\left(E_{S}-E_{0}\right)a(\Delta)\sum_{m,n}2\kappa_{mn}\delta r^{n}\left\langle \uparrow_B\right|S_{1}^{m}\left|\downarrow_B\right\rangle,\label{eq:vso7} \end{equation}
\end{widetext}
where $\left|\uparrow_B\right\rangle$ and $\left|\downarrow_B\right\rangle$ denote the spin states in the direction of the external magnetic field. 
Here $\delta r=\left|\delta\vec{r}\right|$ is the interdot separation, with the displacement vector $\delta\vec{r}$ defined in terms of its components: $\delta r^{n}=r_{R}^{n}-r_{L}^{n}$, where $r_{R(L)}^{n}=\int d^{2}r\left|\chi_{R(L)}\left(\vec{r}\right)\right|^{2}r^{n}$ is the average position of an electron in the right (left) dot. 

Equation (\ref{eq:vso7}) can be further simplified by defining the spin-orbit effective Zeeman field for the translation between the double dot, $\vec{B}_{\rm SO}$, with components
\begin{equation}
g^*\mu_B B_{\rm SO}^{m}= J\sum_{n}\kappa_{mn}\delta r^{n}.\label{eq: HSO3}\end{equation}
The magnitude of $\vec{B}_{\rm SO}$ defines the spin orbit length $l_{\rm SO}$ for translation along the inter-dot direction via the relation $g^*\mu_B\left|\vec{B}_{\rm SO}\right|= J|\delta \vec{r}|/l_{\rm SO}$. 
In addition, note that the eigenvalue equation for 
the matrix $H_S$ in Eq.~(\ref{eq:H_S}), $$H_S \left(\begin{array}{c}a\\ b\end{array}\right) = E_S \left(\begin{array}{c}a\\ b\end{array}\right),$$ yields the identity  $\left(E_{S}-E_{0}\right)a=Jb $. 
Using these two relations in Eq.~(\ref{eq:vso7}), we obtain
\begin{eqnarray}
v_{\rm SO}& =&i\frac{b(\Delta)}{\sqrt{2}}\,2g^*\mu_B\vec{B}_{\rm SO}\cdot\left\langle \uparrow_B\right|\vec{S}_{1}\left|\downarrow_B\right\rangle \label{eq:HSO2}\nonumber\\
&=&\frac{Jb(\Delta)}{\sqrt{2}}\frac{\delta r}{l_{\rm SO}}\left|\vec{\hat{z}}\times\left(\frac{\vec{B}_{\rm SO}}{|\vec{B}_{\rm SO}|}\right)\right|e^{i\varphi_{\rm SO}}.\end{eqnarray}
This leads to Eq.~{\ref{eq:vso}} in this paper, with $\left|\vec{\hat{z}}\times\left(\frac{\vec{B}_{\rm SO}}{|\vec{B}_{\rm SO}|}\right)\right|=\sin{\zeta}$.

The phase $\varphi_{\rm SO}$ depends on the choice of the $x$ and $y$ axes and the direction of the magnetic field. 
In the case when the magnetic field is perpendicular to the 2DEG, we have
 $\left|\vec{\hat{z}}\times\left(\frac{\vec{B}_{\rm SO}}{|\vec{B}_{\rm SO}|}\right)\right|=1$. 
Here one can define the $y$ axis to be in the direction of $\vec{B}_{\rm SO}$, which would make $\varphi_{\rm SO}=0$. 
 In the case the magnetic field has components in the x-y plane, the vector product is no longer necessarily unity. In the special case $\vec{B}\parallel\vec{\vec{B}_{\rm SO}}$, the spin-flip matrix element vanishes. For all other cases the matrix element is not zero. 
Given that the magnetic field direction is defined to be $\vec{z}$, we can always define the x-axis perpendicular to both  $\vec{B}$ and $\vec{\vec{B}_{\rm SO}}$.
This choice again results in $\varphi_{\rm SO}=0$, and hence makes $v_{\rm SO}$ real. 
Thus, in this convention, $v_{\rm SO}$ and the complex matrix element of the transverse Overhauser field, $v_{\rm HF}$, combine like vectors in the $x-y$ plane, with $v_{\rm SO}$ pointing to the x axis, as stated in the main text.
\bibliography{bibdata}

\end{document}